\documentclass[preprint,12pt,sort&compress]{elsarticle}

\usepackage{amssymb}
\usepackage{amsmath}
\usepackage{color}
\usepackage{siunitx}  
\usepackage{orcidlink}  \hypersetup{hidelinks}
\usepackage{cleveref}
\usepackage{longtable}
\usepackage{array}
\usepackage{etoolbox}
\makeatletter 
\patchcmd{\pprintMaketitle}{\itshape\elsaddress\par\vskip36pt}{\itshape\elsaddress\par\vskip18pt}{}{\PackageWarning{main}{title page patch 1 failed}}
\patchcmd{\pprintMaketitle}{\Large\@title\par\vskip18pt}{\Large\@title\par\vskip12pt}{}{\PackageWarning{main}{title page patch 2 failed}}
\patchcmd{\pprintMaketitle}{\printFirstPageNotes}{\printFirstPageNotes\enlargethispage{5\baselineskip}}{}{\PackageWarning{main}{title page patch 3 failed}}
\makeatother

\begin{document}

\begin{frontmatter}

\title{Phase-field simulation of interfacial stability in structured gas-liquid contactors: design rules for gas diffusion electrodes}

\author{Alexander J. Wagner\,\orcidlink{0009-0007-0614-1428}}
\author{Hari Murali Krishna Varma Gottumukkala}
\author{Henning~Bonart\,\orcidlink{0000-0002-5026-4499}\corref{cor1}}
\makeatletter\@eadauthor={Henning Bonart}\makeatother 
\ead{bonart@nmf.tu-darmstadt.de}
\cortext[cor1]{Corresponding author}

\affiliation{organization={Technische Universität Darmstadt, Department of Mechanical Engineering, Institute for Nano- and Microfluidics},
            addressline={Peter-Grünberg-Straße 10}, 
            city={64287 Darmstadt},
            country={Germany}}

\begin{abstract}
Many process engineering applications hold a gas-liquid interface stationary inside a structured solid by a controlled pressure difference.
Gas diffusion electrodes (GDEs) are one such case, where the interface separates the liquid electrolyte from the gaseous reactant.
Maintaining it is particularly challenging in organic electrosynthesis, where low surface tensions lower the admissible pressures throughout and small contact angles leave the electrode with little margin against flooding.
Here, we quantify how the shape, wettability, and three-dimensional topology of a structured medium set this window, using phase-field simulations of the Cahn-Hilliard-Navier-Stokes equations.
Each geometry is characterised by its Laplace pressure curve, which yields both the maximum admissible pressure difference and the stiffness of the interface against pressure fluctuations.
For single flat orifices at constant open area, elongating the opening from an aspect ratio of one to six raises the admissible pressure difference by \SI{40}{\percent} and the stiffness by a factor of $3.3$.
Comparison with slits of equal width shows that this gain originates entirely from the reduced opening width.
Sheared, diamond shaped openings are markedly less effective.
In woven meshes, the wettability displaces the operating window several times as far as it stretches it.
A less wetting surface therefore buys flooding margin largely at the expense of breakthrough margin, whereas the geometric route, established here for flat openings, widens the window itself.
From these results we derive design rules for structured gas-liquid contactors and improved GDE architectures for organic electrosynthesis.
\end{abstract}


\end{frontmatter}

\clearpage

\section{Introduction} \label{sec:introduction}

Many chemical engineering unit operations rely on a gas-liquid interface that is held stationary within a structured or porous solid by a controlled pressure difference, rather than allowed to disperse into bubbles or droplets.
In sieve-plate and bubble-cap distillation and absorption columns, this differential pressure keeps the liquid from draining through the perforations at shutdown or low gas throughput, a failure mode known as weeping \cite{thorat_design_2001, white_drainage_1967}.
In microporous membrane gas-liquid contactors and membrane degassers, the same balance keeps the liquid from breaking through the pores and wetting the membrane, which would otherwise collapse the mass-transfer performance the device is built to provide \cite{mosadegh-sedghi_wetting_2014}.
In each case, the admissible pressure difference and the sensitivity of the interface to fluctuations about it are set jointly by the pore or opening geometry and by the wettability of the solid, yet design practice for these structures remains largely empirical.
Sieve-plate weeping, for instance, is still predicted from device-specific correlations rather than from a general account of how opening shape governs interfacial stability \cite{thorat_design_2001}.

Gas diffusion electrodes (GDEs) present a particularly demanding instance of this same problem and are the motivating application of this work.
A GDE supplies a gaseous reactant directly to a catalyst layer across a gas-liquid interface, bypassing the limited solubility and slow diffusion of gases such as $\mathrm{H_2}$, $\mathrm{O_2}$, $\mathrm{CO_2}$, or $\mathrm{N_2}$ in the surrounding electrolyte, and underlies fuel cells, electrolyzers for $\mathrm{CO_2}$ reduction \cite{baumgartner_narrow_2022}, and, increasingly, organic electrosynthesis, to which the concept is being transferred \cite{wolf_substrate_2025}.
As in the unit operations above, the structured electrode must hold a stable gas-liquid interface under a controlled differential pressure $\Delta p$: below a lower threshold the electrolyte breaches the electrode and floods it, above an upper threshold the gas breaks through and bubbles detach into the electrolyte, and widening this admissible window while keeping the interface stiff against fluctuations is a central design objective that becomes harder to meet as electrodes are scaled to the larger areas required for industrial operation \cite{baumgartner_narrow_2022}.

This challenge is particularly severe in organic electrosynthesis.
Aqueous GDEs benefit from the comparatively high surface tension of water and from hydrophobic polymer or carbon substrates, whose large contact angles give a high water entry pressure and correspondingly wide pressure windows \cite{senocrate_importance_2022}.
Organic electrolytes and non-aqueous reaction media have a substantially lower surface tension, which lowers the admissible pressures throughout, and they remove the hydrophobic repulsion between the liquid and the carbon fibre support on which aqueous GDEs rely, so that the same electrode geometry retains far less margin against flooding \cite{lazouski_non-aqueous_2020}.
Compensating for this loss of chemical wettability contrast through the geometry of the electrode itself, rather than through the substrate chemistry alone, therefore becomes essential in this context, mirroring the role that pore and opening geometry play for the other structures discussed above.

The relevant geometric length scale is set by the openings of the structure, whether they arise from the pores of a membrane or sintered substrate, the perforations of a sieve plate, or the mesh openings of a woven electrode.
Woven meshes are an attractive GDE architecture for this purpose because their weave pattern, wire diameter, and opening shape can be controlled independently of the electrode material.
The experimental evidence on commercial substrates is, however, sobering.
Among the gas diffusion electrodes compared in \cite{baumgartner_narrow_2022}, the woven carbon cloth had the narrowest pressure window.
A wider window was reached only with thicker, denser substrates at the expense of mass transfer, and the one substrate with a markedly wider window, a crack-free non-woven, lost its hydrophobicity during electrolysis.
Baumgartner et al.\ \cite{baumgartner_flooding_2022} propose instead a woven cloth that tolerates continuous electrolyte breakthrough.
Their conclusion, however, is drawn from the commercially available substrates rather than from the capillary mechanism that sets the window.
The pressure window of a woven structure is set by an opening geometry that a weave can be designed to control, and it is this dependence, absent from the commercial materials surveyed there, that the present work quantifies.

At each individual opening, the admissible pressure window follows from the classical Young-Laplace balance between the pressure difference and the curvature the interface can adopt while pinned at the opening \cite{gerlach_quasi-static_2005, lesage_analysis_2013, siedel_toward_2024, oliver_resistance_1977}, which in turn depends on the size and shape of the opening.
Sharp corners likewise arrest sliding droplets by pinning \cite{oliver_resistance_1977, bonart_influence_2019} and affect the stability of liquid films overflowing microstructures \cite{bonart_stability_2020}.
Previous work has related the liquid breakthrough pressure of wire meshes \cite{chhatre_scale_2010} and of perforated micromeshes \cite{acanfora_experimental_2026} to their geometry and wettability, simulated capillary intrusion and drainage in a fuel-cell gas diffusion layer for different wettabilities \cite{sarkezi-selsky_lattice_2022}, and measured both limits of the window for commercial GDEs \cite{baumgartner_narrow_2022}.
To our knowledge, however, how opening shape, wettability, and the three-dimensional topology of a woven structure jointly set both limits of the admissible pressure window and its stiffness has not been resolved systematically for any of these structures, including GDEs.

Quantifying this relationship requires resolving the full family of equilibrium interface shapes an opening can sustain, together with the response of the interface to fluctuations, rather than the breakthrough pressure alone, and because the contact line at the rim of an opening moves as the interface deforms, it requires a numerical method capable of representing moving contact lines without the stress singularity that a conventional no-slip condition otherwise imposes at a sharp interface \cite{huh_hydrodynamic_1971}.
Diffuse-interface, or phase-field, methods resolve this singularity by replacing the sharp interface with a thin transition region governed by a free-energy functional \cite{cahn_free_1958}, regularising the contact-line motion through diffusion \cite{bonart_comparison_2019, jacqmin_contact-line_2000} and thereby avoiding the explicit interface tracking or reconstruction that sharp-interface methods require in geometrically complex, three-dimensional structures \cite{anderson_diffuse-interface_1998, worner_numerical_2012}.
Implementations of this approach have been validated for a range of wetting and bubble dynamics problems relevant to microfluidic and process-engineering applications \cite{cai_numerical_2015, jamshidi_suitability_2019}, making it a suitable tool for the mesh geometries considered here.

In this work, we use phase-field simulations, employing the solver \texttt{phaseFieldFoam} \cite{cai_numerical_2015} in foam-extend \cite{jasak_release_2016}, to quantify how the shape, wettability, and three-dimensional topology of a structured medium set the admissible pressure window of a gas-liquid interface and the stiffness with which it resists displacement, using GDE mesh design for organic electrosynthesis as the motivating application.
We first define, in \cref{sec:pressureCurve}, a set of metrics that characterise this admissible pressure window and the stiffness of the interface within it, and validate our numerical approach against the analytical solution for a circular orifice in \cref{sec:validation}.
Using these metrics, \cref{sec:flatOrifices} isolates the influence of the shape of a single opening on interfacial stability, before \cref{sec:meshes} extends the analysis to three-dimensional, GDE-like woven mesh structures of varying geometry and wettability.
Two results organise what follows.
First, the width of an opening, and not its elongation, sets the admissible pressure difference.
Elongating an opening is worthwhile only as the means of narrowing it at preserved open area.
Second, the wettability of a mesh displaces its operating window much further than it widens it, i.e., a surface treatment trades one failure mode against the other, whereas the geometric route, established here for flat openings, widens the window itself.
From these results, \cref{sec:conclusions} distils design rules for structured gas-liquid contactors more broadly, and applies them to GDE architectures with a wider and more robust operating window, supporting their reliable use in organic electrosynthesis and their scale-up toward industrial application.

\section{Numerical methodology} \label{sec:numericalMethodology}

In this work, interface formation within mesh structures is simulated using a phase-field method.
This section describes the underlying continuous mathematical model, \cref{subsec:CHNS}, the wetting boundary conditions applied at solid walls, \cref{subsec:BCs}, the numerical solver, \cref{subsec:solver}, and the settings that are common to all simulations of this work, \cref{subsec:commonSettings}.
Case-specific domains and boundary conditions are given in the respective sections.
The symbols used throughout this work are collected in \cref{tab:nomenclature}, cf.\ \ref{app:nomenclature}.

\subsection{Governing equations: the Cahn-Hilliard-Navier-Stokes phase-field method} \label{subsec:CHNS}

\subsubsection{Phase-field description of the interface}

To carry out the numerical multiphase simulations, we use a phase-field method based on the convective Cahn-Hilliard-Navier-Stokes equations, a particular type of diffuse-interface models \cite{anderson_diffuse-interface_1998, worner_numerical_2012}.
Unlike sharp-interface methods, such as the volume-of-fluid or level-set methods, which represent the interface as a surface of zero thickness and reconstruct or track it explicitly, diffuse-interface models treat the interface between two immiscible fluids as a transition region of small but finite width, across which physical properties vary steeply but continuously.
While sharp-interface models combined with a no-slip condition exhibit a non-integrable viscous stress singularity at a moving contact line \cite{huh_hydrodynamic_1971}, diffuse-interface models remove this singularity through Cahn-Hilliard diffusion and can therefore represent contact-line motion without a slip model \cite{jacqmin_contact-line_2000, worner_numerical_2012}.
Furthermore, since it is not necessary to explicitly track the interface, all governing equations are solved over the entire computational domain in a single-field formulation, without interface-reconstruction or remeshing algorithms.
These features are essential for the geometrically complex mesh structures considered in \cref{sec:meshes}.

Phase-field models are based on the free energy of the fluid and can be traced back to van der Waals' diffuse-interface theory of capillarity more than a century ago \cite{rowlinson_translation_1979}, in which the free energy was expressed as a functional of both the density and its spatial gradients.
This concept later developed into the Cahn-Hilliard framework \cite{cahn_free_1958}, in which the interface is endowed with an excess free energy that gives rise to its surface tension \cite{jacqmin_calculation_1999}.

The phase distribution of two phases $\mathrm{A}$ and $\mathrm{B}$ is described by a dimensionless order parameter $C$ (the phase field), which represents the difference between the local volumetric phase fractions $\alpha$ of the two phases, i.e., $C=\alpha_\mathrm{A}-\alpha_\mathrm{B}$.
Thus, $C$ takes the distinct values $C_\mathrm{A}=1$ and $C_\mathrm{B}=-1$ in the bulk phases and varies rapidly but smoothly in a thin transition layer, the diffuse interface.
The interface location is conventionally defined by the contour level $C=0$.
In equilibrium, the variation $-0.9\leq C\leq0.9$ occurs over a distance of $4.164\varepsilon$ \cite{jacqmin_calculation_1999}, where $\varepsilon$ is an interfacial thickness parameter, referred to here as the interfacial width.
Through the Cahn number, $Cn=\varepsilon/L_\mathrm{c}$, $\varepsilon$ is related to a characteristic length scale of the macroscopic system, $L_\mathrm{c}$.

The evolution of $C$, and thus of the interface, is governed by the convective Cahn-Hilliard equation
\begin{equation}
    \frac{\partial C}{\partial t} + \left(\mathbf{u}\cdot\nabla\right)C = \kappa\nabla^2\phi \text{,} \label{eq:CH}
\end{equation}
where $t$ denotes time, $\mathbf{u}$ the velocity field obtained from the Navier-Stokes equations described below, and $\kappa$ a constant diffusion parameter, referred to as the mobility.
The chemical potential $\phi$ represents the local variation of the total free energy with respect to $C$.
It comprises two terms, one originating from the homogeneous (bulk-type) part of the free energy, a double-well potential with minima at the bulk values $C=\pm1$, and one from the gradient (interfacial) part,
\begin{equation}
    \phi = \frac{\lambda}{\varepsilon^2}C\left(C^2-1\right) - \lambda\nabla^2C \text{,} \label{eq:phi}
\end{equation}
where $\lambda$ is the mixing energy density.
The surface tension, which is by definition the excess energy present in the system due to the presence of the interface, follows as
\begin{equation}
    \sigma=\frac{2\sqrt{2}}{3}\frac{\lambda}{\varepsilon} \label{eq:sigma}
\end{equation}
\cite{jacqmin_calculation_1999,yue_diffuse-interface_2004}.
In our simulations, we prescribe $\sigma$ and $\varepsilon$ as input parameters, so that $\lambda$ follows from \cref{eq:sigma}.

The right-hand side of \cref{eq:CH} provides a diffusive flux driven by the chemical potential gradient, where the mobility $\kappa$ serves as a numerical parameter controlling the diffusion process.
It is this diffusive mechanism that allows a three-phase contact line to move despite a no-slip condition at walls \cite{jacqmin_contact-line_2000}.

With \cref{eq:phi}, the Cahn-Hilliard \cref{eq:CH} involves fourth-order spatial derivatives of $C$, which makes its numerical treatment complex.
It is therefore retained as a system of the two second-order \cref{eq:CH,eq:phi} for $C$ and $\phi$, cf.\ \ref{app:implementation}.

\subsubsection{Coupled flow equations}

In this paper, two immiscible, incompressible, isothermal Newtonian fluids are considered.
Fluid flow of both these phases is described using a single-field formulation of the Navier-Stokes equations,
\begin{align}
    &\frac{\partial\rho_C\mathbf{u}}{\partial t} + \nabla\cdot\left(\rho_C\mathbf{u}\mathbf{u}\right) = -\nabla p + \nabla\cdot\left\{\mu_C\left[\nabla\mathbf{u}+\left(\nabla\mathbf{u}\right)^T\right]\right\} + \mathbf{f}_\sigma + \rho_C\mathbf{g} \text{,} \label{eq:momentum} \\
    &\nabla\cdot\mathbf{u} = 0 \text{,} \label{eq:continuity}
\end{align}
where $p$ denotes the pressure field and $\mathbf{g}$ the gravitational acceleration.
The density and viscosity fields depend on the order parameter as
\begin{align}
    \rho_C&=\frac{1+C}{2}\rho_\mathrm{A}+\frac{1-C}{2}\rho_\mathrm{B} \text{,} \label{eq:rho} \\ 
    \mu_C&=\frac{1+C}{2}\mu_\mathrm{A}+\frac{1-C}{2}\mu_\mathrm{B} \text{,} \label{eq:mu}
\end{align}
where $\rho_\mathrm{A/B}$ and $\mu_\mathrm{A/B}$ are the density and viscosity of the pure phases.
Following \cite{jacqmin_calculation_1999,villanueva_generic_2006}, the surface tension is incorporated as a volume force 
\begin{equation}
    \mathbf{f}_\sigma=-C\nabla\phi \text{,} \label{eq:f_sigma}
\end{equation}
representing the continuum surface tension force in its potential form.
This formulation converges to the classical (sharp-interface) surface tension force as $\varepsilon\to0$ \cite{lowengrub_quasiincompressible_1998}.

\Cref{eq:rho,eq:mu,eq:f_sigma} couple the Navier-Stokes \cref{eq:momentum} with the Cahn-Hilliard \cref{eq:CH}, which in turn depends on the velocity field $\mathbf{u}$, closing the two-way coupling between the phase and the flow field.

\subsection{Boundary conditions at solid walls} \label{subsec:BCs}

Besides the velocity, the phase-field formulation requires boundary conditions for the order parameter and the chemical potential.
Those imposed at solid walls are given below, whereas those at the remaining boundaries are stated in the respective sections.

The wettability of the solid enters the model through the boundary condition for $C$.
Assuming the interface at the wall to be at or near local equilibrium, the variation of the wall free energy yields the Neumann condition
\begin{equation}
    \mathbf{n}_\mathrm{s}\cdot\nabla C = \frac{\sqrt{2}}{2}\frac{\cos\theta_\mathrm{e}}{\varepsilon}\left(1-C^2\right) \label{eq:wettingBC}
\end{equation}
\cite{villanueva_generic_2006,ding_wetting_2007}, where $\theta_\mathrm{e}$ is the equilibrium contact angle, measured in phase $\mathrm{A}$, and $\mathbf{n}_\mathrm{s}$ is the unit normal pointing from the fluid into the solid.
If the apparent contact angle departs from $\theta_\mathrm{e}$, Cahn-Hilliard diffusion near the wall restores it on a fast time scale \cite{jacqmin_contact-line_2000}.
For contact lines moving far from equilibrium, this surface-energy form no longer reproduces $\theta_\mathrm{e}$ exactly, and a geometric formulation has been proposed instead \cite{ding_wetting_2007}.
For the quasi-static states considered here, the near-equilibrium assumption holds.

For the chemical potential, a homogeneous Neumann condition, $\mathbf{n}_\mathrm{s}\cdot\nabla\phi=0$, is applied, so that no diffusive flux of $C$ crosses the wall and the order parameter is conserved globally \cite{ding_wetting_2007}.
All walls are no-slip, $\mathbf{u}=\mathbf{0}$.

\subsection{Numerical solver} \label{subsec:solver}

The governing equations are solved with the solver \texttt{phaseFieldFoam} \cite{cai_numerical_2015,worner_spreading_2021}, implemented within foam-extend-4.1 \cite{jasak_release_2016}, a community-driven fork of the open-source computational continuum mechanics library OpenFOAM\textsuperscript{\textregistered} \cite{weller_tensorial_1998}.
Further details on the implementation and the employed discretisation schemes are given in \ref{app:implementation}.
The solver has been validated for a range of wetting and bubble dynamics problems \cite{cai_numerical_2015,cai_numerical_2016,fink_drop_2018,worner_spreading_2021,wang_bubble_2022}, and its suitability for gas-liquid microfluidic applications has been demonstrated in comparison with an algebraic volume-of-fluid solver \cite{jamshidi_suitability_2019}.
An additional validation for the configuration of interest here, an interface quasi-statically pushed through a solid structure, is provided in \cref{sec:validation}.

\subsection{Common simulation settings} \label{subsec:commonSettings}

\subsubsection{Fluid properties}

Throughout this manuscript, we consider two immiscible phases, $\mathrm{A}$ and $\mathrm{B}$, with phase $\mathrm{A}$ taken as water and phase $\mathrm{B}$ as air.
Accordingly, we use the common physical properties $\rho_\mathrm{A}=\rho_\mathrm{water}=\SI{998}{\kilo\gram\per\cubic\meter}$, $\mu_\mathrm{A}=\mu_\mathrm{water}=\SI{1}{\milli\pascal\second}$, $\rho_\mathrm{B}=\rho_\mathrm{air}=\SI{1.2}{\kilo\gram\per\cubic\meter}$, $\mu_\mathrm{B}=\mu_\mathrm{air}=\SI{18}{\micro\pascal\second}$, and $\sigma=\SI{72.86}{\milli\newton\per\meter}$.

\subsubsection{Initial conditions}

The phase field is initialised with the analytical equilibrium profile of a planar interface, which is obtained from \cref{eq:CH,eq:phi} for stagnant fluids at steady state, where $\phi\equiv0$,
\begin{equation}
    C(z)=\tanh\left(\frac{z-z_0}{\sqrt{2}\varepsilon}\right) \text{,} \label{eq:C_0}
\end{equation}
representing an interface normal to the domain's $z$-axis at $z=z_0$.
Furthermore, we set $\mathbf{u}=\mathbf{0}$ and $p=0$.

\subsubsection{Spatial and temporal discretisation}

In all simulations, the domain is discretised using uniform, cubical grid cells of size $\Delta x=\Delta y=\Delta z=\ell$.
The complex GDE-like mesh geometries considered in \cref{sec:meshes} are cut out of a block of such cells using \texttt{snappyHexMesh}, which produces skewed cells at the resulting solid surface.
This is accounted for by the numerical schemes listed in \ref{app:implementation}.

The time step width $\Delta t$ is limited by the Courant number, $Co=U_\mathrm{max}\Delta t/\ell\leq Co_\mathrm{max}$, and by an empirically determined upper bound $\Delta t_\mathrm{max}$, so that $\Delta t=\min\left(Co_\mathrm{max}\ell/U_\mathrm{max},\,\Delta t_\mathrm{max}\right)$, where $U_\mathrm{max}$ is the maximum velocity magnitude in the domain.
For the quasi-static cases considered in this paper, $U_\mathrm{max}$ is small, so that $\Delta t=\Delta t_\mathrm{max}$.

The interfacial width $\varepsilon$, the mobility $\kappa$, and the grid size $\ell$ are the remaining free parameters of the method.
They are determined in \cref{subsec:numericalParameters} and kept unchanged throughout the rest of this work.

\section{The Laplace pressure curve as a metric for interfacial stability} \label{sec:pressureCurve}

Throughout this work, the stability of a gas-liquid interface held inside a structure is assessed from the Laplace pressure across the interface as a function of the position of the interface within the opening.
We refer to this relation as the Laplace pressure curve.
Resolving the complete family of equilibrium interface shapes rather than the breakthrough value alone, it yields the largest pressure difference the interface can sustain and the stiffness with which it resists displacement by pressure fluctuations.
Both are needed to compare candidate mesh geometries.

\begin{figure}[ht]
  \centerline{\includegraphics{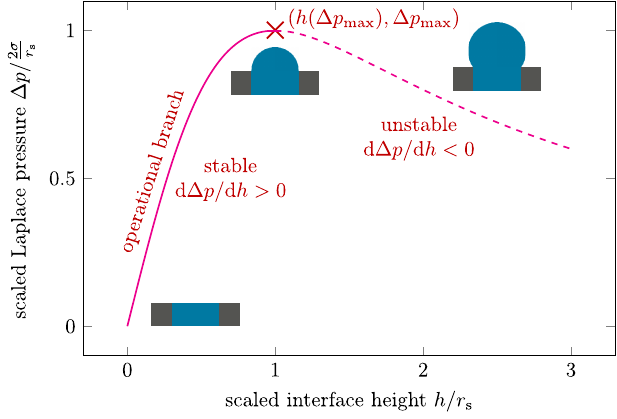}}
  \caption{Laplace pressure curve for a gas bubble quasi-statically pushed through a single circular orifice of radius $r_\mathrm{s}$, obtained from the Young-Laplace equation as described in \cref{subsec:analyticalSolution}. On the operational branch, $\mathrm{d}\Delta p/\mathrm{d}h>0$ and the interface is stable; beyond the maximum $\Delta p_\mathrm{max}$, the equilibrium is unstable and cannot be reached under pressure control.}
\label{fig:pressureCurve}
\end{figure}

\Cref{fig:pressureCurve} shows the curve for the reference configuration of this work, a gas bubble quasi-statically pushed through a single circular orifice of radius $r_\mathrm{s}$ submerged in a liquid, with the three-phase contact line pinned at the orifice rim.
The Laplace pressure $\Delta p$ is the pressure difference across the interface evaluated at the orifice plane, and the interface position is measured by the height $h$ of the highest point of the interface above that plane.
Both are scaled with the corresponding properties of the orifice, $\Delta p$ with its capillary entry pressure $2\sigma/r_\mathrm{s}$ and $h$ with $r_\mathrm{s}$.

Each point of the curve corresponds to an equilibrium interface shape.
Initially, the interface is flat, so that $h=0$ and $\Delta p=0$.
As the gas volume increases, the interface curves and the Laplace pressure increases until it reaches a maximum value $\Delta p_\mathrm{max}\approx2\sigma/r_\mathrm{s}$ at $h\approx r_\mathrm{s}$, where the interface is nearly hemispherical and its apex radius of curvature reaches a minimum, $R_0\approx r_\mathrm{s}$.
For $\mathbf{g}=\mathbf{0}$, this holds exactly, $\Delta p_\mathrm{max}=2\sigma/r_\mathrm{s}$ at $h=r_\mathrm{s}$.
More generally, and for the non-circular openings studied in \cref{sec:flatOrifices,sec:meshes}, $\Delta p_\mathrm{max}=2\sigma/R_\mathrm{min}$, where $R_\mathrm{min}$ is the smallest mean radius of curvature that an interface can attain within the opening.
It is set by the shape of the opening and, unlike for the circular orifice, is generally not equal to any single dimension of it.

For a gas diffusion electrode operated at a controlled Laplace pressure, exceeding $\Delta p_\mathrm{max}$ means the loss of the equilibrium.
The gas breaks through the electrode and bubbles detach into the electrolyte, so that the interface separating the gaseous reactant from the electrolyte, and with it the electrochemical function of the electrode, is lost.
$\Delta p_\mathrm{max}$ is thus the upper limit of the admissible operating window.

Beyond this maximum, the bubble continues to grow while its curvature, and hence $\Delta p$, decreases again.
This descending branch is, however, not accessible if bubble growth is driven by an increasing pressure difference rather than a prescribed inflow.
Thus, the operational branch of a Laplace-pressure-controlled GDE is limited to the ascending branch of the curve.

Within the operational branch, the gradient of the curve measures how strongly the interface resists displacement.
A pressure fluctuation $\delta p$, as it occurs during operation, displaces the interface by $\delta h\approx\delta p\left(\mathrm{d}\Delta p/\mathrm{d}h\right)^{-1}$, so that a steeper branch means a smaller excursion for the same disturbance.

The gradient is, however, not uniform along the branch.
It is largest close to the orifice plane and vanishes at $\Delta p_\mathrm{max}$, so that it has to be evaluated where the electrode is actually operated.
We take as representative operating point the centre of the admissible window, which leaves the same margin against fluctuations in both directions, and define the interface stiffness as the local gradient there,
\begin{equation}
    K = \left.\frac{\mathrm{d}\Delta p}{\mathrm{d}h}\right|_{\Delta p=(\Delta p_\mathrm{max}+\Delta p_\mathrm{min})/2} \text{.} \label{eq:stiffness}
\end{equation}
Here, $\Delta p_\mathrm{min}$ denotes the lower end of the window.
For a pinned flat orifice, the operational branch starts from the flat interface at $\Delta p=0$, so that $\Delta p_\mathrm{min}=0$ and $K$ is evaluated at half the admissible pressure difference, close to how woven GDEs have been run \cite{lazouski_non-aqueous_2020}.
For the woven meshes of \cref{subsec:meshResults}, the window extends to negative pressure differences, and $\Delta p_\mathrm{min}<0$ is the flooding limit.
$K$ is scaled with $(2\sigma/r_\mathrm{s})/r_\mathrm{s}$, consistent with the scaling of $\Delta p$ and $h$.

In the remainder of this work, geometries and operating conditions are therefore compared by three quantities read off their Laplace pressure curves: the maximum admissible pressure difference $\Delta p_\mathrm{max}$, the interface height $h(\Delta p_\mathrm{max})$ at which it is reached, and the interface stiffness $K$, \cref{eq:stiffness}.
\Cref{sec:validation} establishes the accuracy with which the simulations reproduce such a curve, \cref{sec:flatOrifices} uses them to quantify the influence of the shape of a single opening, and \cref{sec:meshes} transfers the results to GDE-like mesh structures with different wetting conditions.

\section{Numerical parameters and validation: bubble growth through a circular orifice} \label{sec:validation}

This section fixes the three remaining parameters of the method: the interfacial width $\varepsilon$, the mobility $\kappa$, and the grid size $\ell$.
Additionally, it verifies that with these values the simulation reproduces a Laplace pressure curve of the kind defined in \cref{sec:pressureCurve}.
The values are then kept unchanged throughout the rest of this work.

The validation case considered here is a gas bubble quasi-statically pushed through a single circular orifice submerged in a liquid, with its contact line pinned at the rim of the orifice.
This case has been studied extensively \cite{gerlach_quasi-static_2005, lesage_analysis_2013, siedel_toward_2024} and can be described theoretically using the Young-Laplace equation, whose predictions have been confirmed against measured bubble profiles for this configuration \cite{gerlach_quasi-static_2005, lesage_analysis_2013}, making it suitable for comparison with simulation results.
It is also the reference geometry against which all openings of \cref{sec:flatOrifices} are scaled.

\subsection{Simulation setup}

We consider a two-dimensional, axisymmetric computational domain, representing an orifice of radius $r_\mathrm{s}=\SI{50}{\micro\meter}$ and height $h_\mathrm{s}=\SI{50}{\micro\meter}$ with a reservoir of radius $r_\mathrm{res}=\SI{100}{\micro\meter}$ and height $h_\mathrm{res}=\SI{150}{\micro\meter}$ on top, cf. \cref{fig:validation}(a).
At the orifice walls, we apply the boundary conditions introduced in \cref{subsec:BCs}, with an equilibrium contact angle of $\theta_\mathrm{e}=\SI{90}{\degree}$ at the side wall of the orifice and $\theta_\mathrm{e}=\SI{30}{\degree}$ at the horizontal wall surrounding the orifice opening.
This ensures that the interface remains flat while moving through the orifice, and only curves and pins once it reaches the orifice rim.

At the inlet, located at $z=0$, we impose a plug-flow inflow condition, $\mathbf{u}=(0,0,U_0)$, together with zero-gradient conditions for the pressure, order parameter, and chemical potential, $\mathbf{n}\cdot\nabla p=0$, $\mathbf{n}\cdot\nabla C=0$, and $\mathbf{n}\cdot\nabla\phi=0$, where $\mathbf{n}$ is the unit vector normal to the boundary.
At the outlet, formed by the top and side boundaries of the reservoir, we prescribe a fixed pressure, $p=0$, together with zero-gradient conditions for the velocity, order parameter, and chemical potential, $\mathbf{n}\cdot\nabla\mathbf{u}=\mathbf{0}$, $\mathbf{n}\cdot\nabla C=0$, and $\mathbf{n}\cdot\nabla\phi=0$.

The Bond number based on the orifice radius, $Bo=(\rho_\mathrm{water}-\rho_\mathrm{air})gr_\mathrm{s}^2/\sigma=\num{3.4e-4}$, is sufficiently small that gravity is marginal at this length scale.
It is nevertheless included as $\mathbf{g}=(0,0,-9.81)\,\si{\meter\per\square\second}$, i.e., directed opposite to the direction of bubble growth, in both the simulation and the reference solution of \cref{subsec:analyticalSolution}.
The interface is initialised as a flat interface at $z_0=\SI{30}{\micro\meter}$ using \cref{eq:C_0}, and the maximum time step limitation is set to $\Delta t_\mathrm{max}=\SI{20}{\nano\second}$.

We choose $U_0=\SI{0.01}{\meter\per\second}$, which keeps the growth quasi-static.
The density and viscosity ratios of this pair, $\rho_\mathrm{A}/\rho_\mathrm{B}=832$ and $\mu_\mathrm{A}/\mu_\mathrm{B}=56$, are large, but the momentum coupling they enter is weak throughout this work.
The inertial and the viscous stress exerted on the interface are negligible against capillarity, $We=\rho_\mathrm{water}U_0^2 r_\mathrm{s}/\sigma=\num{6.8e-5}$ and $Ca=\mu_\mathrm{water}U_0/\sigma=\num{1.4e-4}$, where the properties of the liquid are used because it is the denser and more viscous phase and therefore dominates the resistance to interface motion.

The Laplace pressure curve is recorded in a single run.
Since gas is supplied continuously at $U_0$, the curve is read parametrically, each output time $t$ yielding one point $(h(t),\Delta p(t))$.
Under the quasi-static conditions established above, each of these points is an equilibrium state, so that $\Delta p$ is a function of $h$ alone and the curve is independent of the rate at which the interface is displaced.
The interface relaxes on the inertio-capillary time scale $(\rho_\mathrm{water}r_\mathrm{s}^3/\sigma)^{1/2}\approx\SI{0.04}{\milli\second}$, whereas its apex needs $r_\mathrm{s}/U_0=\SI{5}{\milli\second}$ to advance by $r_\mathrm{s}$.
Because the inflow prescribes the gas volume rather than the pressure, this also holds beyond $\Delta p_\mathrm{max}$, so that the descending branch, which is inaccessible under pressure control (\cref{sec:pressureCurve}), is traced as well.

\subsection{Numerical parameters} \label{subsec:numericalParameters}

Adequately resolving the diffuse interface with the computational grid is critical for obtaining reliable simulation results.
In one dimension and at equilibrium, the width of the diffuse interface in the phase-field method is $L_C=4.164\varepsilon\approx4\varepsilon$ \cite{jacqmin_calculation_1999}.
Thus, the grid resolution of the diffuse interface is quantified by a parameter $N_C=L_C/\ell\approx4\varepsilon/\ell$.
Previous \texttt{phaseFieldFoam} studies found the results to become insensitive to the grid once the diffuse interface is resolved by about six cells or more \cite{cai_numerical_2016, jamshidi_suitability_2019, worner_spreading_2021}, with $N_C=4$--$8$ reported as a good compromise between accuracy and cost \cite{cai_numerical_2015}.
Here, we use $N_C=8$.

In the governing convective Cahn-Hilliard-Navier-Stokes equations presented in \cref{subsec:CHNS}, there are three phase-field-specific parameters: the interfacial width $\varepsilon$, the mobility $\kappa$, and the mixing energy density $\lambda$.
For a given surface tension $\sigma$, $\lambda$ follows from \cref{eq:sigma}, so that only $\varepsilon$ and $\kappa$ still need to be fixed.

The interfacial width $\varepsilon$ is chosen relative to the characteristic length scale of the problem, the orifice diameter $d_\mathrm{s}$.
The Cahn number, $Cn=\varepsilon/d_\mathrm{s}$, quantifies the ratio of the two length scales.
Reported thresholds below which results become largely insensitive to $\varepsilon$ lie between $Cn=0.01$ and $0.02$ and depend on the quantity considered \cite{cai_numerical_2015, cai_numerical_2016, villanueva_generic_2006, jamshidi_suitability_2019,worner_spreading_2021}.
Since decreasing $Cn$ also decreases $\ell$, thereby increasing the total number of grid cells and the computational cost, we use $Cn=0.02$ and verify it for the quantities reported here, cf.\ \cref{subsec:validationResults,subsec:flatSetup}.
For $d_\mathrm{s}=\SI{100}{\micro\meter}$, this yields $\varepsilon=Cn\,d_\mathrm{s}=\SI{2}{\micro\meter}$ and, with $N_C=8$, a grid cell size of $\ell=4\varepsilon/N_C=\SI{1}{\micro\meter}$.

The mobility is scaled with the square of the interfacial width, $\kappa=\chi\varepsilon^2$ with a constant $\chi$ \cite{yue_spontaneous_2007, jamshidi_suitability_2019, worner_spreading_2021}, which corresponds to the lower bound, $O(\varepsilon^2)$, of the mobility scalings between $O(\varepsilon^2)$ and $O(\varepsilon)$ discussed by Jacqmin \cite{jacqmin_calculation_1999}.
We use $\chi=\SI{1}{\meter\second\per\kilogram}$, which for $\varepsilon=\SI{2}{\micro\meter}$ yields $\kappa=\SI{4e-12}{\cubic\meter\second\per\kilogram}$.

The double-well potential underlying the chemical potential, \cref{eq:phi}, admits no stationary solution that conserves the phase volume at finite curvature \cite{dadvand_advected_2021}, so that a curved interface loses volume by bulk diffusion.
The amount lost grows with $Cn$, and the rate at which it is lost grows with $\kappa$ \cite{yue_spontaneous_2007, jamshidi_suitability_2019}.
However, the results reported here are not obtained from the gas volume.
The interface is displaced by a prescribed inflow and the Laplace pressure is read against the simultaneously measured interface height, so that a loss of gas shifts the instant at which a given height is reached but not the pair $(h,\Delta p)$ itself.
Its residual effect on the equilibrium shapes is bounded by the agreement with the analytical solution in \cref{sec:validation} and by the interface refinement in \cref{subsec:flatSetup}, both \SI{0.3}{\percent} in $\Delta p_\mathrm{max}$.

\subsection{Analytical reference solution} \label{subsec:analyticalSolution}

Under quasi-static conditions, viscous and inertial stresses are negligible, and the interface shape is governed by the Young-Laplace equation, which balances the local pressure jump across the interface with the surface tension acting on its mean curvature \cite{gerlach_quasi-static_2005, lesage_analysis_2013, siedel_toward_2024},
\begin{equation}
  \Delta p=\sigma\left(\frac{1}{R_1}+\frac{1}{R_2}\right) \text{.} \label{eq:YoungLaplace}
\end{equation}
Here, $R_1$ and $R_2$ denote the principal radii of curvature.
We describe the axisymmetric interface in a coordinate system with its origin at the bubble apex, where $\tilde{x}$ is the radial distance from the symmetry axis and $\tilde{z}$ is the depth below the apex.
Accounting for the hydrostatic pressure variation, the pressure jump at depth $\tilde{z}$ is
\begin{equation}
    \Delta p(\tilde{z}) = \frac{2\sigma}{R_0} - \left(\rho_\mathrm{A}-\rho_\mathrm{B}\right)g\tilde{z} \text{,} \label{eq:dp_of_z}
\end{equation}
where $R_0$ is the radius of curvature at the apex, where both principal radii coincide by symmetry.

Expressing the principal radii of curvature in terms of the interface profile $\tilde{z}(\tilde{x})$ and combining \cref{eq:YoungLaplace} and \cref{eq:dp_of_z} gives an ordinary differential equation for the interface profile for a prescribed $R_0$, which has no closed-form solution and must therefore be integrated numerically \cite{gerlach_quasi-static_2005}.
Written for $\tilde{z}(\tilde{x})$, it becomes singular where the interface tangent is vertical, i.e., once the bubble extends beyond the hemispherical shape.
Rather than interchanging the dependent and the independent variable along the profile \cite{gerlach_quasi-static_2005, lesage_analysis_2013}, we parametrise the profile by the arc length $s$, measured from the apex, and the angle $\varphi$ between the interface tangent and the horizontal.
This yields the regular first-order system
\begin{align}
    \frac{\mathrm{d}\tilde{x}}{\mathrm{d}s} &= \cos\varphi \text{,} &
    \frac{\mathrm{d}\tilde{z}}{\mathrm{d}s} &= \sin\varphi \text{,} &
    \frac{\mathrm{d}\varphi}{\mathrm{d}s} &= \frac{2}{R_0}-\frac{\left(\rho_\mathrm{A}-\rho_\mathrm{B}\right)g\tilde{z}}{\sigma}-\frac{\sin\varphi}{\tilde{x}} \text{,} \label{eq:arcLength}
\end{align}
with the initial conditions $\tilde{x}=\tilde{z}=\varphi=0$ at $s=0$, where the azimuthal curvature has the removable limit $\sin\varphi/\tilde{x}\to1/R_0$.
 
For a prescribed apex radius of curvature, the system, \cref{eq:arcLength}, is integrated numerically.
Each profile that reaches the orifice radius, $\tilde{x}=r_\mathrm{s}$, intersects it twice, once before and once after its widest point, and therefore contributes one shape to the ascending branch and one to the descending branch of the Laplace pressure curve.
The depth $\tilde{z}$ of the crossing gives the corresponding apex height $h$ above the orifice plane.
Sweeping $R_0$ generates the complete family of equilibrium shapes.
Finally, evaluating \cref{eq:dp_of_z} at the orifice plane, $\tilde{z}=h$, gives the Laplace pressure,
\begin{equation}
    \Delta p_\mathrm{s} = \frac{2\sigma}{R_0}-\left(\rho_\mathrm{A}-\rho_\mathrm{B}\right)gh \text{,} \label{eq:dp_orifice}
\end{equation}
yielding the analytical Laplace pressure curve.

\subsection{Results} \label{subsec:validationResults}

\begin{figure}[!ht]
  \centerline{\includegraphics{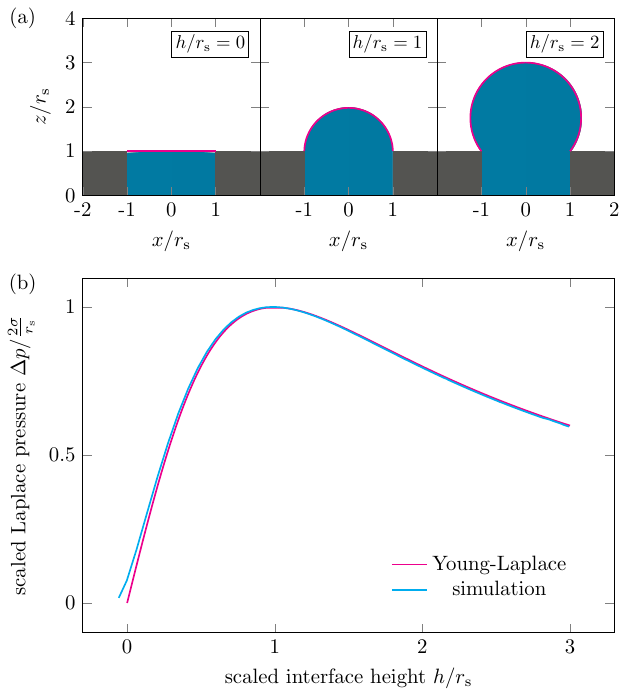}}
  \caption{Comparison of the analytical Young-Laplace solution (magenta) and simulation (blue) for quasi-static gas bubble growth through a circular orifice of radius $r_\mathrm{s}$. (a) Interface profiles at different apex heights $h$. The axisymmetric simulation is mirrored about the symmetry axis, grey denotes the solid, white the liquid and blue the gas. (b) Laplace pressure curves.}
\label{fig:validation}
\end{figure}

\Cref{fig:validation}(a) compares the analytically predicted and simulated bubble profiles for different apex heights $h$, while \cref{fig:validation}(b) shows the corresponding Laplace pressure curves.
The profiles are indistinguishable at plotting resolution.
The simulated Laplace pressure curve reproduces the maximum of the reference solution to within \SI{0.3}{\percent} and the interface height at which it occurs to within \SI{2}{\percent}.
The only noticeable discrepancy occurs for small interface heights $h$.
This deviation arises because the simulation represents the interface as a diffuse interface of finite thickness, whereas the analytical solution assumes a sharp interface.
As the diffuse interface approaches the orifice edge, it begins to curve before the interface midpoint, $C=0$, which defines the interface position, reaches the orifice plane.
Consequently, the interface is already slightly curved at $h=0$, resulting in a small positive Laplace pressure instead of zero.

Over the operational branch above $h=0.25\,r_\mathrm{s}$, the two curves differ by at most \SI{2.6}{\percent} of $\Delta p_\mathrm{max}$.
Since the deviation flattens the lower part of the curve, the simulated interface stiffness, $K=1.58$, is slightly lower than the analytical value of $1.62$.
Although this discrepancy could be reduced by decreasing the interfacial width $\varepsilon$, doing so would increase the computational cost, and this small deviation is considered acceptable.
The chosen numerical parameters ($N_C=8$, $Cn=0.02$, $\chi=\SI{1}{\meter\second\per\kilogram}$) are therefore adequate, and the three metrics that \cref{sec:pressureCurve} defines for comparing geometries---$\Delta p_\mathrm{max}$, $h(\Delta p_\mathrm{max})$, and the interface stiffness $K$---are resolved to a few percent, which is well below the differences between geometries reported in \cref{sec:flatOrifices}.

\section{Interfaces in single, flat orifices} \label{sec:flatOrifices}

After validating the method and selecting suitable numerical parameters, this section studies interfaces in single, flat orifices.
Compared to the woven GDE-like structures studied in \cref{sec:meshes}, single flat orifices can be studied with a simpler setup and provide more generalisable results.
They isolate the effect of the shape of the opening, which is the quantity a mesh manufacturer can actually control, from the three-dimensional topography of a woven structure.
They also isolate it from the wettability of the mesh, which is varied in \cref{sec:meshes}.
Throughout this section, the contact line is pinned at the orifice rim over the whole range of interface heights considered.

We study quasi-static bubble growth through rectangular orifices with different aspect ratios, \cref{subsec:rectangles}, compare them with slits of the same width in order to separate the influence of the width of the opening from that of its length, \cref{subsec:slits}, and consider diamond shaped orifices with different opening angles, \cref{subsec:diamonds}.
Throughout this section, all openings are compared at the same open area, namely that of a circular orifice of radius $r_\mathrm{s}=\SI{50}{\micro\meter}$, $A_\mathrm{s}=\pi r_\mathrm{s}^2=\SI{7854}{\square\micro\meter}$.
The open area is the natural constraint for this comparison because it fixes the area available for gas diffusion into the electrolyte phase at the electrode, so that geometries compared at constant $A_\mathrm{s}$ are compared at equal gas supply capability.
This equivalent radius $r_\mathrm{s}$ is also used to scale the Laplace pressure curves, so that a given curve can be read as the performance of an opening relative to a circular orifice of the same open area.

With gravity negligible at this length scale ($Bo=3.4\times10^{-4}$, cf.\ \cref{sec:validation}), every equilibrium shape is a surface of constant mean curvature pinned at the orifice rim.
Its shape, and hence $h(\Delta p_\mathrm{max})$, is set by the geometry of the opening alone, and $\Delta p_\mathrm{max}$ is proportional to $\sigma$.
In the scaled variables used here, the curves are therefore independent of the surface tension, and the results obtained with the properties of \cref{subsec:commonSettings} apply unchanged to the surface tensions of organic electrolytes, where the same curve simply corresponds to smaller absolute pressures.

That these equilibria are surfaces of constant mean curvature does not mean that solving the Young-Laplace equation alone would have sufficed.
The merging of neighbouring interfaces found in \cref{sec:meshes} is a change of interface topology that a static computation of an equilibrium surface cannot follow.
Computing the pinned cases in the same framework is what makes the comparison across \cref{sec:flatOrifices,sec:meshes} a like-for-like one, while the analytical solution of \cref{subsec:analyticalSolution} serves to verify it.

\subsection{Simulation setup} \label{subsec:flatSetup}

The simulation setup in this section is largely the same as in \cref{sec:validation}, with the orifice geometry being varied, requiring a three-dimensional domain in Cartesian coordinates.
Orifices of height $h_\mathrm{s}=\SI{50}{\micro\meter}$ are again topped by reservoirs of respectively adapted dimensions, and the inlet, outlet, and wall boundary conditions of \cref{sec:validation} are retained.
The reservoirs extend at least $1.5\,x_\mathrm{s}$ laterally and $\SI{100}{\micro\meter}$ in height, which was verified to be large enough that the interface does not interact with the outlet boundary.
Making use of the two symmetry planes of all geometries considered here, only one quarter of the domains is simulated, with symmetry boundary conditions applied at the two symmetry planes.
The resulting meshes comprise between $0.53\times10^6$ and $1.03\times10^6$ cells.

The inflow velocity is again set to $U_0=\SI{0.01}{\meter\per\second}$, ensuring quasi-static conditions, and gravity is included as $\mathbf{g}=(0,0,-9.81)\,\si{\meter\per\square\second}$.
The equilibrium contact angle is $\theta_\mathrm{e}=\SI{90}{\degree}$ at the side walls of the orifice and $\theta_\mathrm{e}=\SI{30}{\degree}$ at the horizontal wall surrounding the orifice opening, so that the interface remains flat inside the orifice and pins at the orifice rim.
The interface is initialised as a flat interface sitting inside the orifice at $z_0=\SI{30}{\micro\meter}$, and the maximum time step limitation is set to $\Delta t_\mathrm{max}=\SI{10}{\nano\second}$.
The numerical parameters determined in \cref{sec:validation} are retained, i.e., $\varepsilon=\SI{2}{\micro\meter}$, $\ell=\SI{1}{\micro\meter}$, and $\chi=\SI{1}{\meter\second\per\kilogram}$.
The Cahn number is thereby defined with respect to the equivalent diameter $d_\mathrm{s}=2r_\mathrm{s}$, which is kept constant for all geometries considered.
Since the narrowest openings studied below are considerably smaller than $d_\mathrm{s}$, we verified that this remains adequate by repeating the simulation of the narrowest slit ($x_\mathrm{s}=\SI{36.2}{\micro\meter}$, corresponding to $AR=6$, cf. \cref{subsec:slits}) with $\varepsilon$, $\ell$, and $\kappa$ scaled down to $Cn=0.02$ with respect to $x_\mathrm{s}$, a refinement of the interface by a factor of $2.8$.
$\Delta p_\mathrm{max}$ changes by \SI{0.3}{\percent}, and the two curves deviate by less than \SI{1}{\percent} of $\Delta p_\mathrm{max}$ for $h\geq0.25\,r_\mathrm{s}$.
The remaining difference is confined to small interface heights and reflects the diffuse-interface offset discussed in \cref{sec:validation}, which decreases with decreasing $\varepsilon$ as expected.

As introduced in \cref{sec:pressureCurve}, the interface position is measured by the height $h$ of the interface apex above the orifice plane, and the Laplace pressure $\Delta p$ is evaluated as the pressure difference across the interface at the orifice plane.

\subsection{Rectangular orifices with different aspect ratios} \label{subsec:rectangles}

To start with, we consider rectangular orifices with different aspect ratios $AR=y_\mathrm{s}/x_\mathrm{s}\in\{1,\allowbreak2,\allowbreak3,\allowbreak4,\allowbreak5,\allowbreak6\}$, where $x_\mathrm{s}$ and $y_\mathrm{s}$ are the orifice width and length, respectively.
With the open area fixed at $A_\mathrm{s}=x_\mathrm{s}y_\mathrm{s}=\pi r_\mathrm{s}^2$, the two side lengths follow as $x_\mathrm{s}=r_\mathrm{s}\sqrt{\pi/AR}$ and $y_\mathrm{s}=r_\mathrm{s}\sqrt{\pi AR}$.
The resulting orifice dimensions are listed in \cref{tab:rectangles}.
Plane, rectangular orifices coincide with the top view of a woven GDE mesh, so that the results obtained here can be used to assess the influence of the aspect ratio of the mesh openings on interfacial stability with a simplified, generalisable approach.

\begin{figure}[ht]
  \centerline{\includegraphics{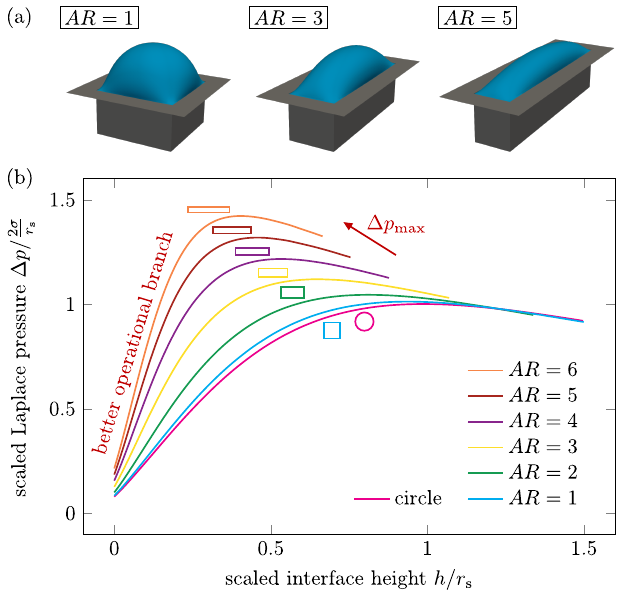}}
  \caption{Quasi-static bubble growth through rectangular orifices with different aspect ratios $AR$ at constant open area. (a) Interface shapes for selected aspect ratios at $h=h(\Delta p_\mathrm{max})$. (b) Laplace pressure curves, together with the curve of the area-equivalent circular orifice.}
\label{fig:rectangles}
\end{figure}

\Cref{fig:rectangles}(a) shows the interface shapes for selected aspect ratios at $h=h(\Delta p_\mathrm{max})$, where the Laplace pressure reaches its maximum.
\Cref{fig:rectangles}(b) shows the corresponding Laplace pressure curves, and \cref{tab:rectangles} lists the three metrics of \cref{sec:pressureCurve} read off them.
The interface is nearly hemispherical for $AR=1$, but becomes increasingly elongated with increasing aspect ratio.
At the same time, the interface height at which the Laplace pressure reaches its maximum decreases.
The pinned contact line forces the interface to curve much more strongly across the short dimension of the opening than along the long one.

\begin{table}[ht]
\centering
\caption{Rectangular orifices of aspect ratio $AR$ at constant open area $A_\mathrm{s}=\pi r_\mathrm{s}^2$, and the area-equivalent circular orifice, characterised by the three metrics of \cref{sec:pressureCurve}. Pressures are scaled with $2\sigma/r_\mathrm{s}$, interface heights with $r_\mathrm{s}$, and $K$ with $(2\sigma/r_\mathrm{s})/r_\mathrm{s}$.}
\label{tab:rectangles}
\begin{tabular}{lccccccc}
\hline
 & \multicolumn{6}{c}{rectangle} & circle \\
$AR$ & 1 & 2 & 3 & 4 & 5 & 6 & --- \\
\hline
$x_\mathrm{s}/\si{\micro\meter}$                      & 88.6  & 62.7  & 51.2  & 44.3  & 39.6  & 36.2  & --- \\
$y_\mathrm{s}/\si{\micro\meter}$                      & 88.6  & 125.3 & 153.5 & 177.2 & 198.2 & 217.1 & --- \\
\hline
$\Delta p_\mathrm{max}$                               & 1.01 & 1.05 & 1.12 & 1.22 & 1.32 & 1.42 & 1.00 \\
$h(\Delta p_\mathrm{max})$                            & 0.94 & 0.81 & 0.65 & 0.53 & 0.46 & 0.40 & 0.99 \\
$K$                                                   & 1.70  & 2.12  & 2.89  & 3.80  & 4.70  & 5.61  & 1.58 \\
\hline
\end{tabular}
\end{table}

The consequences are visible in the Laplace pressure curves.
The maximum Laplace pressure increases monotonically with the aspect ratio, from $\Delta p_\mathrm{max}=1.01\,(2\sigma/r_\mathrm{s})$ at $AR=1$ to $1.42\,(2\sigma/r_\mathrm{s})$ at $AR=6$, so that the interface can withstand a $\SI{40}{\percent}$ larger pressure difference before detachment occurs.
Additionally, the maximum is reached at a considerably smaller interface height, $h(\Delta p_\mathrm{max})$ decreasing from $0.94\,r_\mathrm{s}$ to $0.40\,r_\mathrm{s}$.
The operational branch therefore becomes markedly steeper.
The interface stiffness $K$, \cref{eq:stiffness}, increases by a factor of $3.3$ between $AR=1$ and $AR=6$, meaning that a given pressure fluctuation displaces the interface by only about a third as much.
Both effects act in the same direction, favouring elongated openings.

\Cref{fig:rectangles}(b) also includes the Laplace pressure curve of the area-equivalent circular orifice.
It is very similar to that of the square orifice.
The two maxima differ by only $\SI{1.2}{\percent}$ and the corresponding interface heights by $\SI{5}{\percent}$.
For nearly isotropic openings, the admissible pressure difference is thus governed by the open area alone, and the precise shape of the opening is irrelevant.
It shows that the gains reported above are genuinely an effect of anisotropy, and it justifies using the circular orifice of \cref{sec:validation} as the baseline against which all other geometries are scaled.

These results suggest that GDE meshes with elongated openings offer a wider and more robust operating window than meshes with nearly isotropic openings of the same open area.
The comparison in \cref{subsec:slits} shows which geometric feature of the opening is responsible for this.

\subsection{Rectangular orifices compared to slits} \label{subsec:slits}

In \cref{subsec:rectangles}, the aspect ratio was varied at constant open area, so that an increase in $AR$ is necessarily accompanied by a decrease in the orifice width $x_\mathrm{s}$.
Both the narrower opening and the finite length of the orifice contribute to the resulting Laplace pressure curves.
To separate the two effects, we compare each rectangle with a slit of the same width $x_\mathrm{s}$ and infinite length, which is the limiting case $AR\to\infty$ at fixed width.
Slits are translationally invariant along their length, so that they can be simulated in a two-dimensional domain that is otherwise identical to the setup described in \cref{subsec:flatSetup}.

For a slit, only one principal curvature remains.
With gravity being negligible at the length scale considered here ($Bo\approx3.4\times10^{-4}$, cf. \cref{sec:validation}), \cref{eq:YoungLaplace} is then solved by a circular arc pinned at the two rims, and the Laplace pressure attains its maximum for the semi-cylindrical interface, $h=x_\mathrm{s}/2$, where the arc radius equals the half-width of the slit,
\begin{equation}
    \Delta p_\mathrm{max,slit} = \frac{\sigma}{x_\mathrm{s}/2} = \frac{2\sigma}{x_\mathrm{s}} \text{.} \label{eq:slitMax}
\end{equation}

\begin{figure}[ht]
  \centerline{\includegraphics{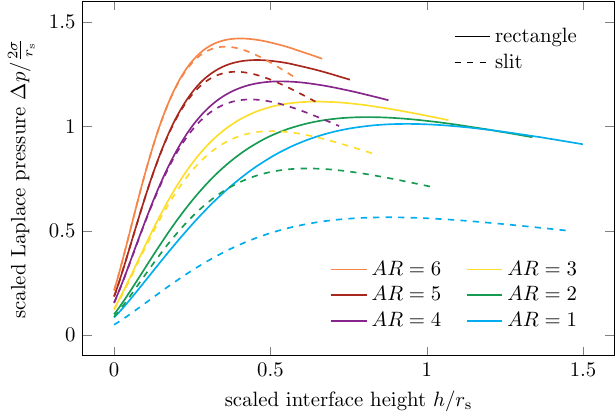}}
  \caption{Laplace pressure curves for quasi-static bubble growth through rectangular orifices of different aspect ratios $AR$ compared to corresponding slits of width $x_\mathrm{s}$ and infinite length.}
\label{fig:rectanglesVsSlits}
\end{figure}

\Cref{fig:rectanglesVsSlits} compares the Laplace pressure curves of the rectangles with those of the corresponding slits.
For every aspect ratio, the rectangle lies above its slit, i.e., the finite length of the opening always stabilises the interface.
The additional pressure is the contribution of the second principal curvature, which the interface acquires at the two short ends of the rectangle and which is absent in the slit.
Writing the maximum pressure of the rectangle as
\begin{equation}
    \Delta p_\mathrm{max,rect} = f(AR)\,\frac{2\sigma}{x_\mathrm{s}} \label{eq:endCapFactor}
\end{equation}
isolates this contribution in the factor $f\geq1$, which is listed in \cref{tab:rectsVsSlits} and decays rapidly with increasing aspect ratio.
For $AR\geq4$, the slit reproduces not only $\Delta p_\mathrm{max,rect}$ to within \SI{8}{\percent}, but the complete ascending branch of the pressure curve, whereas for the square orifice it underestimates the admissible pressure difference by almost a factor of two.

\begin{table}[ht]
\centering
\caption{Maximum Laplace pressure of the rectangular orifices and of the corresponding slits of width $x_\mathrm{s}=r_\mathrm{s}\sqrt{\pi/AR}$, the end-cap factor $f=\Delta p_\mathrm{max,rect}/\Delta p_\mathrm{max,slit}$ of \cref{eq:endCapFactor}, and the interface stiffness $K$, \cref{eq:stiffness}, for both geometries.
Pressures are scaled with $2\sigma/r_\mathrm{s}$, and $K$ with $(2\sigma/r_\mathrm{s})/r_\mathrm{s}$.}
\label{tab:rectsVsSlits}
\begin{tabular}{lcccccc}
\hline
$AR$ & 1 & 2 & 3 & 4 & 5 & 6 \\
\hline
$x_\mathrm{s}/\si{\micro\meter}$    & 88.6 & 62.7 & 51.2 & 44.3 & 39.6 & 36.2 \\
$\Delta p_\mathrm{max,rect}$        & 1.01 & 1.05 & 1.12 & 1.22 & 1.32 & 1.42 \\
$\Delta p_\mathrm{max,slit}$        & 0.56 & 0.80 & 0.98 & 1.13 & 1.26 & 1.38 \\
$f$                                 & 1.79 & 1.31 & 1.15 & 1.08 & 1.04 & 1.03 \\
\hline
$K$, rectangle                      & 1.70 & 2.12 & 2.89 & 3.80 & 4.70 & 5.61 \\
$K$, slit                           & 1.00 & 1.97 & 2.89 & 3.84 & 4.76 & 5.66 \\
\hline
\end{tabular}
\end{table}

This decomposition also explains the trend reported in \cref{subsec:rectangles}.
Going from $AR=1$ to $AR=6$ at constant open area narrows the opening by a factor $\sqrt{6}$ and thus raises $2\sigma/x_\mathrm{s}$ by a factor of $2.45$, while the end-cap factor $f$ drops from $1.79$ to $1.03$.
The two effects act against each other, and the resulting net gain in $\Delta p_\mathrm{max}$ is only a factor of $1.40$.
The benefit of elongated openings therefore originates entirely from their reduced width; the elongation itself removes a stabilising contribution that a nearly isotropic opening still possesses.
Elongation is thus not a mechanism of stabilisation but the means by which the opening can be narrowed while the open area, and with it the transport performance of the mesh, is preserved.

Consequently, the simple estimate
\begin{equation}
    \Delta p_\mathrm{max}\approx\frac{2\sigma}{x_\mathrm{s}} \label{eq:designRule}
\end{equation}
is useful for mesh design, but only as long as the opening is strongly elongated, i.e., as long as its smallest dimension is substantially smaller than its largest one.
For $AR\geq4$ it underestimates $\Delta p_\mathrm{max}$ by less than \SI{8}{\percent}, while it fails for nearly isotropic openings.
For the square orifice, both principal curvatures contribute equally, so that $\Delta p_\mathrm{max}$ approaches the value $2\sigma/r_\mathrm{s}$ of the area-equivalent circular orifice, and \cref{eq:designRule} underestimates the admissible pressure difference by almost a factor of $2$.
Since the slit provides a lower bound for $\Delta p_\mathrm{max}$ for every shape considered here, inexpensive two-dimensional slit approximations can nevertheless be used for a conservative first screening of mesh geometries, and are quantitatively accurate for the elongated openings that are of interest for GDE design.

For $AR\geq3$, the slit is also accurate for the stiffness.
Its $K$ agrees with that of the rectangle to within \SI{1.5}{\percent}, cf.\ \cref{tab:rectsVsSlits}.
For $AR\geq4$ it is even slightly higher, since its $\Delta p_\mathrm{max}$ is lower, so that $K$ is evaluated at a lower pressure and hence on a lower, steeper part of the curve.
The end caps thus lose their influence on the stiffness faster than on $\Delta p_\mathrm{max}$.
At $AR=1$, they raise $K$ above that of the corresponding slit by \SI{70}{\percent} and at $AR=2$ by \SI{7}{\percent}, whereas from $AR=3$ on they leave it unchanged although they still raise $\Delta p_\mathrm{max}$ by \SI{15}{\percent}.

\subsection{Diamond shaped orifices with different opening angles} \label{subsec:diamonds}

The rectangular openings considered so far have parallel sides, so that their width is constant along the opening.
Woven meshes, however, do not necessarily produce rectangular openings.
By an adapted weaving method or by shearing the mesh during fabrication, the openings can be made rhombic, with a width that varies along the opening.
We therefore consider diamond shaped orifices with opening angle $\alpha$, again at the constant open area $A_\mathrm{s}=\pi r_\mathrm{s}^2$.

Denoting by $x_\mathrm{s}$ and $y_\mathrm{s}$ the short and the long diagonal of the diamond, i.e., the dimensions of its bounding box, the open area is $A_\mathrm{s}=x_\mathrm{s}y_\mathrm{s}/2$ and the two diagonals follow as $x_\mathrm{s}=r_\mathrm{s}\sqrt{2\pi\tan(\alpha/2)}$ and $y_\mathrm{s}=r_\mathrm{s}\sqrt{2\pi/\tan(\alpha/2)}$.
The aspect ratio of the bounding box, $AR=y_\mathrm{s}/x_\mathrm{s}=\cot(\alpha/2)$, is defined exactly as for the rectangles and allows a direct comparison of the two families of shapes at the same elongation.
We vary the opening angle over $\alpha\in\{\SI{30}{\degree},\allowbreak\SI{60}{\degree},\allowbreak\SI{90}{\degree}\}$, corresponding to $AR=3.73$, $1.73$, and $1$.
The case $\alpha=\SI{90}{\degree}$ is the square orifice of \cref{subsec:rectangles} rotated by \SI{45}{\degree}.
We therefore re-use the $AR=1$ result and use it as link between the two series.
Note that keeping the open area constant while reducing $\alpha$ requires the side length of the diamond to grow from $\SI{88.6}{\micro\meter}$ at $\alpha=\SI{90}{\degree}$ to $\SI{125.3}{\micro\meter}$ at $\alpha=\SI{30}{\degree}$, i.e., a sheared mesh has to be woven with a correspondingly coarser pitch to retain the open area of the square weave.

\begin{figure}[ht]
  \centerline{\includegraphics{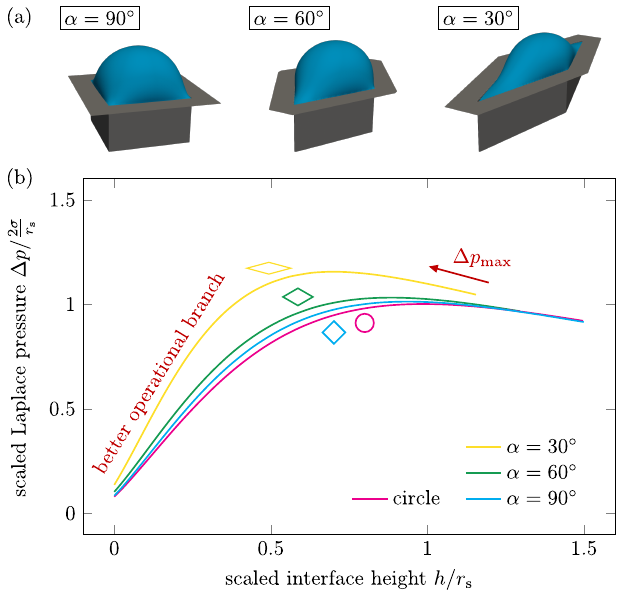}}
  \caption{Quasi-static bubble growth through diamond shaped orifices with different opening angles $\alpha$ at constant open area. (a) Interface shapes at $h=h(\Delta p_\mathrm{max})$. (b) Laplace pressure curves, together with the curve of the area-equivalent circular orifice.}
\label{fig:diamonds}
\end{figure}

\Cref{fig:diamonds}(a) shows the interface shapes at $h=h(\Delta p_\mathrm{max})$ and \cref{fig:diamonds}(b) the corresponding Laplace pressure curves.
Qualitatively, the behaviour matches that of the rectangles.
As the opening angle is reduced, the interface becomes elongated along the long diagonal, the maximum Laplace pressure increases, and it is reached at a smaller interface height.
Quantitatively, however, the gain is considerably weaker.
Reducing $\alpha$ from \SI{90}{\degree} to \SI{60}{\degree} raises $\Delta p_\mathrm{max}$ from $1.01$ to only $1.03$ in units of $2\sigma/r_\mathrm{s}$, even though the bounding box is already elongated by a factor of $1.73$.
Only the sharpest opening, $\alpha=\SI{30}{\degree}$, gives a noticeable improvement of $\Delta p_\mathrm{max}=1.16$ at $h(\Delta p_\mathrm{max})=0.70\,r_\mathrm{s}$.

\begin{table}[ht]
\centering
\caption{Diamond shaped orifices compared with rectangular orifices of the same bounding-box aspect ratio $AR$ and the same open area.
Pressures are scaled with $2\sigma/r_\mathrm{s}$, and $K$ with $(2\sigma/r_\mathrm{s})/r_\mathrm{s}$.
Since $\cot(\alpha/2)$ is not an integer, the rectangle values are linearly interpolated between the simulated aspect ratios of \cref{subsec:rectangles}.}
\label{tab:diamonds}
\begin{tabular}{lccc}
\hline
$\alpha$ & \SI{90}{\degree} & \SI{60}{\degree} & \SI{30}{\degree} \\
\hline
$x_\mathrm{s}$/\si{\micro\meter}                                          & 125.3 & 95.2  & 64.9  \\
$y_\mathrm{s}$/\si{\micro\meter}                                          & 125.3 & 164.9 & 242.1 \\
$AR=\cot(\alpha/2)$                                                       & 1.00  & 1.73  & 3.73  \\
\hline
$\Delta p_\mathrm{max}$, diamond                                          & 1.01 & 1.03 & 1.16 \\
$\Delta p_\mathrm{max}$, rectangle at same $AR$                           & 1.01 & 1.04 & 1.19 \\
\hline
$K$, diamond                                                              & 1.70  & 1.82  & 2.60  \\
$K$, rectangle at same $AR$                                               & 1.70  & 2.01  & 3.55  \\
\hline
\end{tabular}
\end{table}

\Cref{tab:diamonds} compares the diamonds with rectangles of the same bounding-box aspect ratio and the same open area.
At $AR=1.73$ the two shapes are almost indistinguishable, and at $AR=3.73$ the maximum of the diamond falls short of that of the rectangle by only \SI{3}{\percent}.
The difference in the operational branch is much more pronounced.
The stiffness of the diamond is \SI{27}{\percent} lower than that of the equally elongated rectangle, $K=2.60$ instead of $3.55$, and at $AR=1.73$ it is already \SI{9}{\percent} lower.
For interfacial stability, which depends on both quantities, elongating an opening by sharpening two of its corners is therefore clearly less effective than elongating it at constant width.

The reason lies in the area a diamond occupies within its bounding box.
A rhombus fills only half of it, so that at equal open area and equal $AR$ its largest width exceeds the width of the corresponding rectangle by a factor of $\sqrt{2}$, e.g., $\SI{64.9}{\micro\meter}$ instead of $\SI{45.9}{\micro\meter}$ at $AR=3.73$.
Since the Laplace pressure is uniform along the interface, it is this largest width that limits $\Delta p_\mathrm{max}$.
That the diamond nevertheless almost reaches the maximum of the rectangle is due to its converging rims, which impose additional curvature along the entire opening.
Relative to $2\sigma/x_\mathrm{s}$, the diamond gains a factor of $1.50$, the rectangle only $1.09$.
The two effects very nearly cancel in $\Delta p_\mathrm{max}$, but they do not cancel on the operational branch.
There, the interface is only weakly curved, and its response is dominated by the wide central part of the opening.
This is why the diamond has the flatter operational branch and reaches its maximum at a larger interface height, $h=0.70\,r_\mathrm{s}$ instead of $0.56\,r_\mathrm{s}$ for the rectangle, interpolated from \cref{tab:rectangles}.
 
The estimate of \cref{eq:designRule} therefore cannot be transferred to tapered openings.
Inserting the largest width $x_\mathrm{s}$ of the diamond gives $0.77$ and inserting its mean width $x_\mathrm{s}/2$ gives $1.54$, whereas the simulation yields $1.16$.
It is the parallel sides of a rectangle that make its width a single, meaningful design parameter.

Both routes away from the standard square weave therefore improve interfacial stability, but they are not equally efficient.
Shearing the mesh to $\alpha=\SI{30}{\degree}$ raises $\Delta p_\mathrm{max}$ by \SI{14}{\percent} and the stiffness $K$ by \SI{53}{\percent} relative to the square, whereas weaving rectangular openings of the same open area and the same elongation achieves \SI{18}{\percent} and \SI{109}{\percent}.
If a deviation from the standard square weave is made at all, elongating the opening at constant width is thus the more rewarding one.

In principle, the two routes can also be combined into a sheared, elongated opening.
The results above suggest, however, that little is to be gained from this.
A shear along the long axis of an elongated opening leaves the perpendicular distance between its two long sides unchanged if its open area is kept constant, and by \cref{eq:designRule} it is this distance that governs $\Delta p_\mathrm{max}$.
Only the two ends of the opening are modified, and their contribution is already small for elongated openings, $f=1.03$ at $AR=6$.

\section{Interfaces in GDE-like mesh structures} \label{sec:meshes}

\Cref{sec:flatOrifices} isolated the influence of the shape of a single opening on interfacial stability by replacing the electrode with an orifice of the same open area in a flat plate.
An essential feature of a real woven electrode was thereby suppressed.
Its wires are smooth and convex, so that the pinning invoked in \cref{sec:validation,sec:flatOrifices} does not apply and the contact line is free to slide along the wire surface so that the prescribed equilibrium contact angle is attained.
The admissible pressure window therefore ceases to be a property of the geometry alone and becomes a function of the wettability of the electrode.
Additionally, a mesh has a finite thickness, so that the interface apex can reside inside the structure as well as above it.
The interface can consequently be displaced downwards into the mesh by a negative pressure difference, a branch of the Laplace pressure curve that a flat orifice does not possess and that ends with flooding of the electrode.
 
This section addresses these features.
We simulate quasi-static interface displacement through a GDE-like woven mesh of circular wires for three equilibrium contact angles, $\theta_\mathrm{e}\in\{\SI{30}{\degree},\allowbreak\SI{50}{\degree},\allowbreak\SI{70}{\degree}\}$, measured in the liquid as defined in \cref{subsec:BCs}.
The mesh has the same open area as the geometries in \cref{sec:flatOrifices}, and we compare its Laplace pressure curves with that of the area-equivalent flat square orifice of \cref{subsec:rectangles}.

\subsection{Simulation setup} \label{subsec:meshSetup}
 
\begin{figure}[ht]
  \centerline{\includegraphics{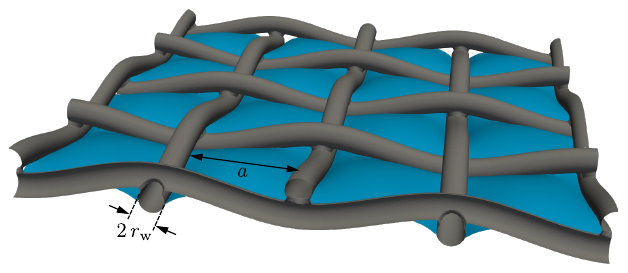}}
  \caption{Section of $4\times4$ openings of the modelled plain square weave with the gas-liquid interface (blue) for $\theta_\mathrm{e}=\SI{30}{\degree}$ at $\Delta p=0.5\,\Delta p_\mathrm{max}$.
  The gas is below the mesh and the liquid above it.
  The wire diameter $2r_\mathrm{w}$ and the side $a$ of the projected square opening are indicated.
  The section was obtained by mirroring the simulated single opening at its symmetry planes.}
\label{fig:meshGeometry}
\end{figure}

The mesh is modelled as a plain square weave of two orthogonal families of circular wires of radius $r_\mathrm{w}=\SI{10}{\micro\meter}$, cf. \cref{fig:meshGeometry}.
Each wire follows a cubic spline that passes alternately above and below the wires it crosses, reproducing the over-under topology of a woven fabric, and the two families intersect at an angle of \SI{90}{\degree}.
At every crossing point, the two wires are joined by a weld of size $r_\mathrm{w}/2=\SI{5}{\micro\meter}$, removing the geometric singularity of two tangentially touching cylinders.

The wire pitch $w$, i.e., the distance between the centrelines of two neighbouring parallel wires, is chosen such that the open area of one mesh opening seen in top view equals the reference area of \cref{sec:flatOrifices}, $A_\mathrm{s}=\pi r_\mathrm{s}^2=\SI{7854}{\square\micro\meter}$ with $r_\mathrm{s}=\SI{50}{\micro\meter}$.
The projected opening is a square of side $a=w-2r_\mathrm{w}$, so that $a=\SI{88.6}{\micro\meter}$, which is exactly the square orifice of \cref{subsec:rectangles}, and $w=\SI{108.6}{\micro\meter}$.
Comparing the mesh with the flat square orifice at equal projected open area isolates the effect of the three-dimensional topography and of the wettability from that of the opening size, in the same way that all shapes of \cref{sec:flatOrifices} were compared at equal $A_\mathrm{s}$.
 
The mesh is positioned in the domain such that its mid-plane lies at $z=0$.
Making use of the symmetry of the weave, a single opening is simulated, with symmetry boundary conditions applied at the four lateral boundaries at $x=0$, $x=w$, $y=0$, and $y=w$, so that the domain represents an infinitely extended mesh.
Reservoirs are attached below and above the mesh, and their extent is adapted to the direction in which the interface is displaced, cf.\ below.
For the interface pushed upwards, the domain extends from $z_\mathrm{res,min}=\SI{-30}{\micro\meter}$ to $z_\mathrm{res,max}=\SI{70}{\micro\meter}$ and comprises $\num{1.10e6}$ cells.
For the interface pushed downwards, it extends from $z_\mathrm{res,min}=\SI{-40}{\micro\meter}$ to $z_\mathrm{res,max}=\SI{30}{\micro\meter}$ and comprises $\num{0.75e6}$ cells.
In each case, the reservoir is extended only on the side into which the interface travels.
The interface does not interact with the inlet or outlet boundary over the range of heights reported below.
The wetting boundary condition of \cref{subsec:BCs} is applied at the entire wire surface with a single equilibrium contact angle $\theta_\mathrm{e}$, which is varied between the three cases. 
Unlike in \cref{sec:validation,sec:flatOrifices}, no auxiliary second contact angle is used, since the mesh possesses no edge at which the contact line is to be pinned.
The numerical parameters determined in \cref{sec:validation} are retained, $\varepsilon=\SI{2}{\micro\meter}$, $\ell=\SI{1}{\micro\meter}$, and $\chi=\SI{1}{\meter\second\per\kilogram}$.
The maximum time step limitation is set to $\Delta t_\mathrm{max}=\SI{10}{\nano\second}$, as in \cref{subsec:flatSetup}.
The geometry is cut out of the uniform Cartesian background grid with \texttt{snappyHexMesh}, cf.\ \cref{subsec:commonSettings}, and the skewness and non-orthogonality corrections listed in \ref{appsec:schemes} are active for these grids.
 
The Laplace pressure curve is assembled from two runs that start from the same relaxed state.
In both, the interface is initialised as a flat interface at $z_0=0$ using \cref{eq:C_0}, with $p=0$ prescribed at the bottom and the top boundary, and is allowed to relax to a steady state.
The resulting equilibrium defines the reference height $h_0$, i.e., the position the interface adopts in the mesh when the electrode is operated at zero differential pressure.
From this state, the first run imposes an inflow $U_0=\SI{0.01}{\meter\per\second}$ at the bottom boundary and pushes the interface upwards through and out of the mesh, generating the branch $h>h_0$.
The second run imposes $U_0=\SI{-0.01}{\meter\per\second}$ at the top boundary and pushes the interface downwards through and out of the mesh, generating the branch $h<h_0$.
The two branches are joined at $(h_0,\,0)$ to give the complete curve.
 
The interface height $h$ is measured at the centre of the opening, $x=y=w/2$, as the height of the $C=0$ contour above the mid-plane of the mesh.
The Laplace pressure $\Delta p$ is evaluated as the difference between the gas pressure below the mesh and the static pressure of the electrolyte.
As throughout \cref{sec:flatOrifices}, $\Delta p$ is scaled with $2\sigma/r_\mathrm{s}$ and $h$ with $r_\mathrm{s}$, so that the curves can be read directly against those of the area-equivalent flat orifices.

\subsection{Results} \label{subsec:meshResults}
 
\begin{figure}[ht]
  \centerline{\includegraphics{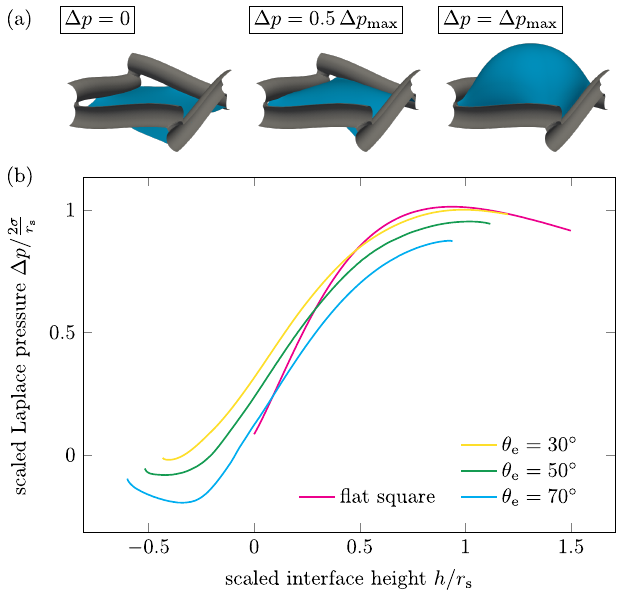}}
  \caption{Quasi-static displacement of a gas-liquid interface through mesh structures of different equilibrium contact angles $\theta_\mathrm{e}$. (a) Interface shapes for the intermediate contact angle $\theta_\mathrm{e}=\SI{50}{\degree}$ and different Laplace pressures. (b) Laplace pressure curves, together with the curve of the area-equivalent flat square orifice.}
\label{fig:wovenMeshes}
\end{figure}

\Cref{fig:wovenMeshes}(a) shows the interface in the mesh with $\theta_\mathrm{e}=\SI{50}{\degree}$ at three differential pressures.
At $\Delta p=0$, the interface is almost flat and lies inside the weave.
At $\Delta p=0.5\,\Delta p_\mathrm{max}$, it has risen to about $0.2\,r_\mathrm{s}$ above the mid-plane of the mesh and is only weakly curved.
The apex of the interface is still located below the wire crests.
Finally, at $\Delta p=\Delta p_\mathrm{max}$, it forms a dome protruding well above the wire crests, with the contact line now resting on the upper flanks of the wires.
Over the whole sequence, the contact line is not fixed but migrates along the wire surface, and the interface meets the wires at the prescribed angle $\theta_\mathrm{e}$ throughout.
This is an essential difference from \cref{sec:flatOrifices}, where the contact line was held at the orifice rim over the entire range of interface heights, and the interface adjusted only its curvature.
Here, it adjusts its curvature and the position of its own boundary.

\Cref{fig:wovenMeshes}(b) shows the corresponding Laplace pressure curves for the three contact angles, together with that of the area-equivalent flat square orifice of \cref{subsec:rectangles}, and \cref{tab:meshes} lists the metrics read off them.
In \cref{fig:wovenMeshes}(b), the ends of the Laplace pressure curves are not chosen but set by the geometry.
Each curve terminates where the interfaces of neighbouring openings come into contact and fully merge across the wires that separate them.
Beyond this point, the simulation passes into a dynamic transition in which the entire mesh dewets, for the interface pushed upwards, or floods, for the interface pushed downwards.
The merging height therefore marks the end of the family of admissible equilibrium shapes, and it moves closer to the mid-plane as the contact angle increases, because the contact line advances further across the wire crests before the interface has grown to a given height.
On the liquid side, the corresponding limits all lie beyond $h(\Delta p_\mathrm{min})$, so that the minima reported below are attained; the margin is smallest for $\theta_\mathrm{e}=\SI{30}{\degree}$.
 
On the gas side, this is not the case for $\theta_\mathrm{e}=\SI{70}{\degree}$.
Here, the merging limit falls very close to the maximum, so that the last, nearly flat part of the stable branch is cut off before the maximum is properly resolved.
The maximum pressure $\Delta p_\mathrm{max}$ the interface could sustain is therefore reduced.
The stiffness $K$ is evaluated well below the maximum and is therefore resolved, but it inherits a weak dependence on the truncation through $\Delta p_\mathrm{max}$.

\begin{table}[ht]
\centering
\caption{Metrics of GDE-like woven mesh structures of different equilibrium contact angles $\theta_\mathrm{e}$, compared with the area-equivalent flat square orifice of \cref{subsec:rectangles}.
$h_0$ is the interface height at $\Delta p=0$.
Pressures are scaled with $2\sigma/r_\mathrm{s}$, interface heights with $r_\mathrm{s}$, and $K$ with $(2\sigma/r_\mathrm{s})/r_\mathrm{s}$.
Heights are measured from the mid-plane of the mesh.
Values marked $^{\dagger}$ are affected by the merging of neighbouring interfaces before the maximum is fully resolved, cf.\ \cref{subsec:meshResults}.}
\label{tab:meshes}
\begin{tabular}{lcccc}
\hline
 & \multicolumn{3}{c}{woven mesh} & flat square \\
$\theta_\mathrm{e}$ & \SI{30}{\degree} & \SI{50}{\degree} & \SI{70}{\degree} & --- \\
\hline
$h_0$                                                        & $-0.33$ & $-0.20$ & $-0.09$ & $0$ \\
\hline
$\Delta p_\mathrm{max}$                                      & $1.00$ & $0.95$ & $0.87^{\dagger}$ & $1.01$ \\
$h(\Delta p_\mathrm{max})$                                   & $0.99$ & $1.02$ & $0.92^{\dagger}$ & $0.94$ \\
$K$                                                          & $1.25$ & $1.25$ & $1.31^{\dagger}$ & $1.70$ \\
\hline
$\Delta p_\mathrm{min}$                                      & $-0.02$ & $-0.08$ & $-0.20$ & --- \\
$h(\Delta p_\mathrm{min})$                                   & $-0.40$ & $-0.42$ & $-0.34$ & --- \\
\hline
$\Delta p_\mathrm{w}$                                        & $1.02$ & $1.04$ & $1.07^{\dagger}$ & --- \\
\hline
\end{tabular}
\end{table}

From \cref{fig:wovenMeshes} and \cref{tab:meshes}, three observations follow.
First, for all three contact angles, the equilibrium height is negative, $h_0/r_\mathrm{s}=-0.33$, $-0.20$, and $-0.09$ for $\theta_\mathrm{e}=\SI{30}{\degree}$, \SI{50}{\degree}, and \SI{70}{\degree}, so that the interface recedes further down into the weave the better the electrolyte wets the wires.
This follows from the wetting boundary condition together with the convexity of the wires.
At $\Delta p=0$, the interface is flat and would meet a vertical wall at \SI{90}{\degree}.
To meet the wire at a smaller angle, it must move to a position at which the wire surface is correspondingly inclined, which on a circular wire lies below its widest section.
The more wetting the surface, the further down the wire the contact line has to sit, and the deeper the interface is drawn into the structure.

Second, unlike the curves of \cref{sec:flatOrifices}, those of \cref{fig:wovenMeshes}(b) extend to negative pressures and exhibit a minimum $\Delta p_\mathrm{min}<0$ in addition to the maximum $\Delta p_\mathrm{max}$.
The two extrema bound the operating window from both sides and correspond to the two failure modes of a GDE described in the introduction.
Exceeding $\Delta p_\mathrm{max}$ detaches the gas into the electrolyte, while falling below $\Delta p_\mathrm{min}$ results in electrode flooding.
Between the two, the branch $\mathrm{d}\Delta p/\mathrm{d}h>0$ forms the operational range of the electrode.
The set of metrics defined in \cref{sec:pressureCurve} therefore has to be extended for mesh structures by a non-zero $\Delta p_\mathrm{min}$ and by the width of the resulting window,
\begin{equation}
    \Delta p_\mathrm{w} = \Delta p_\mathrm{max} - \Delta p_\mathrm{min} \text{.} \label{eq:windowWidth}
\end{equation}
Since the window now has two ends, \cref{eq:stiffness} is evaluated with the actual, negative $\Delta p_\mathrm{min}$, i.e., at the operating point that keeps the same margin against both failure modes.

Third, increasing the contact angle from \SI{30}{\degree} to \SI{70}{\degree} lowers $\Delta p_\mathrm{max}$ from $1.00$ to $0.87$ in units of $2\sigma/r_\mathrm{s}$, while it lowers $\Delta p_\mathrm{min}$ from $-0.02$ to $-0.20$.
Thus, between \SI{30}{\degree} and \SI{50}{\degree}, where all four extrema are resolved, the width of the window, \cref{eq:windowWidth}, changes only from $1.02$ to $1.04$.
At \SI{70}{\degree}, $\Delta p_\mathrm{max}$ is affected by the merging of neighbouring interfaces, and so is $\Delta p_\mathrm{w}=1.07$.
Nevertheless, due to the decrease in $\Delta p_\mathrm{min}$, the window is about \SI{5}{\percent} wider than at \SI{30}{\degree}.
Within the range studied here, the wettability of the mesh therefore governs where the admissible pressure window lies relative to $\Delta p=0$ far more than how wide it is.

This is seen most directly in the curves themselves.
For $h>h_0$, the three curves of \cref{fig:wovenMeshes}(b) run almost parallel.
They are separated by a vertical offset of about $0.07$ and $0.09\,(2\sigma/r_\mathrm{s})$ on average between $\theta_\mathrm{e}=\SI{30}{\degree}$ and \SI{50}{\degree} and between \SI{50}{\degree} and \SI{70}{\degree}, which decreases slightly towards the maximum.
The contact angle thus translates the pressure curve without appreciably changing its shape, and the stiffness with which the interface resists a given displacement is, on this branch, essentially a property of the geometry alone, $K=1.25$ at both \SI{30}{\degree} and \SI{50}{\degree}.
The slightly larger value at \SI{70}{\degree}, $K=1.31$, follows from the lower window midpoint of that curve rather than from a stiffer interface.
The three curves differ in shape only for $h<h_0$, where they approach their respective minima at different rates.
That branch is, however, the one the electrode is not operated on, since it lies below the pressure-balanced state and terminates in flooding, so the difference there does not affect the design conclusions drawn from the operational branch.

Measurements on porous GDE substrates show the same division of roles between wettability and geometry.
On carbon-based GDEs, a catalyst layer that lowers the contact angle shifts the pressure window towards larger gas overpressures, as it does here, without significantly changing its width \cite{baumgartner_narrow_2022}.
On a PTFE substrate of \SI{0.3}{\micro\meter} pore size, a surface treatment that renders the polymer wettable lowers the water entry pressure from \SI{9.2e5}{\pascal} to practically zero \cite{senocrate_importance_2022}.

\subsection{Comparison with the flat orifice} \label{subsec:meshVsFlat}
 
For the most wetting mesh, $\theta_\mathrm{e}=\SI{30}{\degree}$, the maximum admissible pressure difference is $\Delta p_\mathrm{max}=1.00\,(2\sigma/r_\mathrm{s})$, nearly matching the $1.01\,(2\sigma/r_\mathrm{s})$ obtained for a flat square orifice with the same projected open area.
The three-dimensional topography of the weave, the curvature of the wires, and the welds at the crossings therefore cost essentially nothing in terms of the gas-side limit, and the flat, area-equivalent opening predicts $\Delta p_\mathrm{max}$ of a well-wetted mesh to within a few percent.
The reduction of $\Delta p_\mathrm{max}$ at larger contact angles, by \SI{6}{\percent} at \SI{50}{\degree} and by at most \SI{14}{\percent} at \SI{70}{\degree}, is not solely a topographic effect but mainly the price of the missing pinning edge.
In \cref{sec:flatOrifices}, the sharp rim allowed the apparent contact angle to vary freely over a finite range, so that the interface could attain its minimum radius of curvature irrespective of $\theta_\mathrm{e}$, while on a smooth wire it cannot, and the attainable curvature is limited by the wetting condition.

A second consequence of the free contact line acts in the same direction.
Because the wires are round, the aperture that the interface has to span is not the projected opening but the clear width between the wires at the height at which the contact line actually sits.
The projected side length $a$ is the width at the widest section of the wires; above it the wire surface curves away from the opening, so that a contact line resting above that section spans a larger distance $a_\mathrm{eff}$.
By the same argument that fixes $h_0$, a larger contact angle places the contact line higher on the wire flanks, so that $a_\mathrm{eff}$ at a given interface height grows with $\theta_\mathrm{e}$.
With the design rule of \cref{eq:designRule}, $\Delta p_\mathrm{max}\approx2\sigma/a_\mathrm{eff}$, a wider effective aperture directly means a lower admissible pressure difference.
The relevant width of a woven mesh is therefore not the width of its openings seen from above, but the width seen at the level of the three-phase contact line.
This also explains why the flat, area-equivalent orifice reproduces the mesh so well at $\theta_\mathrm{e}=\SI{30}{\degree}$ and progressively worse as $\theta_\mathrm{e}$ increases.

The stiffness of the operational branch, in contrast to $\Delta p_\mathrm{max}$, is not recovered by the flat orifice.
With $K=1.25$ at $\theta_\mathrm{e}=\SI{30}{\degree}$ and \SI{50}{\degree}, the mesh is about a quarter softer than the flat square orifice, $K=1.70$.
This, too, is a consequence of the free contact line.
On the flat orifice, the interface can accommodate an increase in pressure only by curving more strongly, whereas in the mesh it can also move its boundary along the wires, cf.\ \cref{subsec:meshResults}, so that the same pressure increment displaces it further.
 
The merging that truncates the curve is, however, not a limitation of the model but a property of the mesh.
Since it occurs at $\theta_\mathrm{e}=\SI{70}{\degree}$ before the interface has reached its pressure maximum, the mesh at this contact angle does not fail by the mechanism assumed throughout \cref{sec:pressureCurve,sec:flatOrifices}, in which the equilibrium of an isolated opening is lost, but by coalescence of the interfaces of adjacent openings, which the single-opening analysis of \cref{sec:flatOrifices} cannot describe at all.
The operating limit of the electrode is then the merging pressure rather than $\Delta p_\mathrm{max}$.
Determining this merging pressure requires simulations of woven mesh structures, as performed here, rather than single-opening analyses.
Its onset, on the other hand, is resolved only to within the interface width.
Two diffuse interfaces merge once their profiles overlap, i.e., once they approach within a few $\varepsilon$, whereas physically, the liquid film between them must first drain to molecular thickness before it ruptures.
The computed merging pressure is therefore a lower bound and would rise somewhat with decreasing $\varepsilon$.
 
Two design consequences follow for GDEs operated in organic electrolytes, where $\theta_\mathrm{e}$ is small.
First, a well-wetted mesh sustains almost the full capillary entry pressure of its projected opening on the gas side, but has virtually no margin below $\Delta p=0$.
The electrode must therefore be operated at a distinctly positive differential pressure rather than near the pressure-balanced state that the geometry adopts on its own, and the control system has to guarantee this offset under all fluctuations of the electrolyte pressure.
The steel cloth of Lazouski et al.\ \cite{lazouski_non-aqueous_2020} behaved in this way.
A well-defined separation of gas and electrolyte was obtained at non-zero pressure differences, and the electrode was operated at \SI{1000}{\pascal}, about half its breakthrough pressure.
Second, since the wettability displaces the window much further than it widens it, a treatment that lowers the wettability buys flooding margin largely at the expense of breakthrough margin.
Widening the window itself requires the geometric route of \cref{sec:flatOrifices}: reducing the smallest dimension of the opening at constant open area.
In a weave, however, narrowing the opening may cause the merging limit to be reached before $\Delta p_\mathrm{max}$.
Testing this route on a weave requires varying the pitch and the wire radius, which are held fixed here.

Throughout this section, one equilibrium contact angle is prescribed over the entire wire surface, whereas a real electrode carries contact angle hysteresis, chemical heterogeneity and roughness.
All three pin the contact line to some degree.
A real mesh should therefore fall between the freely sliding contact line assumed here and the fully pinned behaviour of \cref{sec:flatOrifices}, so that the two sections bracket it rather than describing it separately.

\section{Conclusions and outlook} \label{sec:conclusions}

We used phase-field simulations of the Cahn-Hilliard-Navier-Stokes equations to quantify how the shape, the wettability, and the three-dimensional topology of a structured medium set the pressure window over which it holds a stationary gas-liquid interface.
Rather than the breakthrough pressure alone, we resolved the complete family of equilibrium shapes an opening sustains.
Three metrics are read off the resulting Laplace pressure curve, cf.\ \cref{sec:pressureCurve}: the maximum admissible pressure difference, the interface excursion needed to reach it, and the interface stiffness, which is defined as the local gradient of the operational branch at the centre of the admissible window.
The last of these fixes how far a pressure fluctuation displaces the interface and cannot be obtained from a breakthrough measurement.
For a circular orifice, the method reproduces the Young-Laplace solution to within \SI{0.3}{\percent} in $\Delta p_\mathrm{max}$, cf.\ \cref{sec:validation}.

\subsection{Design rules} \label{subsec:designRules}

For structured gas-liquid contactors generally, whether the openings are the perforations of a sieve plate \cite{thorat_design_2001, white_drainage_1967}, the pores of a membrane contactor \cite{mosadegh-sedghi_wetting_2014}, or the openings of a woven electrode, our results give the following.

\begin{enumerate}

\item \emph{The smallest width $x_\mathrm{s}$ of the opening sets the admissible pressure difference.}
For elongated openings with parallel sides, $\Delta p_\mathrm{max}\approx2\sigma/x_\mathrm{s}$, \cref{eq:designRule}, to better than \SI{8}{\percent} for $AR\geq4$ and conservatively throughout, cf.\ \cref{subsec:slits}.
Where the openings of an electrode are not identical, by design or through manufacturing tolerances, breakthrough sets in at the opening with the largest $x_\mathrm{s}$, since all of them are held at the same gas pressure.

\item \emph{Elongation narrows the opening but does not itself stabilise the interface.}
On the contrary, it removes the stabilising end-cap contribution that a nearly isotropic opening still possesses, cf.\ \cref{subsec:slits}.
Its value lies in being the one route to a narrower opening at preserved open area, and hence at preserved gas supply.
At constant area, going from $AR=1$ to $AR=6$ gains \SI{40}{\percent} in $\Delta p_\mathrm{max}$ and a factor of $3.3$ in stiffness.

\item \emph{Parallel sides make the width a meaningful design parameter; tapered openings do not.}
A rhombus fills only half of its bounding box, so its largest width exceeds that of the equivalent rectangle by $\sqrt{2}$.
Its converging rims nearly compensate this in $\Delta p_\mathrm{max}$ but not on the operational branch, which leaves the stiffness \SI{27}{\percent} lower, and \cref{eq:designRule} cannot be transferred to such openings, cf.\ \cref{subsec:diamonds}.

\item \emph{In a mesh of round wires, the relevant width is the clear width at the height of the contact line.}
The contact line slides freely, so a larger contact angle places it higher on the wire flanks, where the effective aperture exceeds the projected opening.
The area-equivalent flat orifice therefore predicts the positive admissible pressure for a well-wetted mesh very accurately and does progressively worse as $\theta_\mathrm{e}$ grows, cf.\ \cref{subsec:meshVsFlat}.
It does not predict the stiffness, which the free contact line lowers by about a quarter.

\item \emph{Wettability locates the operating pressure window, geometry sizes it.}
Over the contact angles resolved here, the window is displaced several times as far as it is stretched, cf.\ \cref{subsec:meshResults}.
A less wetting surface therefore buys flooding margin largely at the expense of breakthrough margin, whereas rules 1 and 2, established for flat openings, widen the window itself.
In organic electrolytes, where $\theta_\mathrm{e}$ is small, the binding constraint is flooding and not breakthrough, and the electrode has to be held at a distinctly positive differential pressure.

\item \emph{Two-dimensional slits screen candidate geometries conservatively.}
The slit bounds $\Delta p_\mathrm{max}$ from below for every shape considered and reproduces the stiffness of rectangles with $AR\geq3$ to within \SI{1.5}{\percent}, cf.\ \cref{subsec:slits}.

\end{enumerate}

Gravity is negligible at this length scale, so every equilibrium shape is a surface of constant mean curvature.
In the scaled variables used here, the curves are therefore independent of the surface tension.
The rules carry over unchanged to the lower surface tensions of organic electrolytes, where the same curves simply correspond to smaller absolute pressures.
The chemistry enters through $\theta_\mathrm{e}$ alone.

These rules widen the operating window rather than making the electrode tolerant of leaving it.
For aqueous $\mathrm{CO_2}$ electrolysis, Baumgartner et al.\ \cite{baumgartner_flooding_2022} argue the opposite case, that the windows of commercially available substrates are too narrow to scale and that a material which keeps working while flooded is the more practical route.
The two positions are complementary rather than opposed.
Their argument rests on the pore structure of given commercial materials, whereas rules 1 and 2 apply to a weave whose opening geometry is itself a design variable.
Their remedy, moreover, relies on a bimodal pore structure that drains the electrolyte through large pores while smaller ones stay gas filled, which a single-scale metallic weave does not possess.
It also requires the smaller pores to repel the electrolyte, which is least likely in organic electrolytes, where the present rules matter most.
In organic electrolytes the flooding margin is already the binding constraint, so tolerating electrolyte breakthrough spends the margin that is scarce.

\subsection{Outlook} \label{subsec:outlook}

The rules above were obtained for a single opening at a time, and the most immediate extension is to give up that restriction.
\Cref{sec:meshes} already shows a failure mode that a single opening cannot exhibit, in which the interfaces of neighbouring openings merge across the wire between them before either reaches its own pressure maximum.
Mapping where in the space of pitch, wire radius and contact angle this limit takes over from $\Delta p_\mathrm{max}$ would complete the picture the present work begins.

Experimental validation against woven electrodes in organic electrolytes remains a necessary next step, since our results rest on verification against analytical solutions rather than against measured pressure windows.
Beyond that, three directions seem worth following.
In an operating cell, the electrolyte flows past the interface, which exposes the part protruding into the channel to shear and to a non-uniform dynamic pressure.
Because the Cahn-Hilliard-Navier-Stokes formulation solves the full momentum equation, such a crossflow could be included without any change to the method, so that the admissible Laplace pressure and the stiffness can be mapped against the Weber number of the flow.
Coupling the interface mechanics to the electrochemistry would resolve gas consumption and reaction-driven pressure transients, and thus the fluctuations against which the stiffness of the operational branch is the relevant defence.
A chemically patterned or graded wettability would escape the trade-off that rule 5 imposes on a uniform surface, and could relocate the operating window without spending breakthrough margin to buy flooding margin.

\section*{CRediT authorship contribution statement}

\textbf{Alexander J. Wagner:} Data curation, Formal analysis, Investigation, Methodology, Software, Validation, Visualization, Writing -- original draft, Writing -- review \& editing.

\textbf{Hari Murali Krishna Varma Gottumukkala:} Investigation, Validation.

\textbf{Henning Bonart:} Conceptualization, Funding acquisition, Methodology, Project administration, Resources, Supervision, Validation, Writing -- original draft, Writing -- review \& editing.

\section*{Declaration of competing interest}

The authors declare that they have no known competing financial interests or personal relationships that could have appeared to influence the work reported in this paper.

\section*{Data availability}

Data will be made available on request.

\section*{Acknowledgements}

We thank Bastian Etzold, Philipp Röse, Nils Näser, and Hans-Joachim Kohnke for helpful discussions during the project.

This work was funded by the German Federal Ministry of Research, Technology and Space (BMFTR) within the Clusters4Future cluster ETOS through the project GDE4OES (grant no.\ 03ZU1205MA).

The authors gratefully acknowledge the computing time provided to them at the NHR Center NHR4CES at TU Darmstadt (project numbers 2620 and 26428).
This is funded by the Federal Ministry of Research, Technology and Space, and the state governments participating on the basis of the resolutions of the GWK for national high performance computing at universities (www.nhr-verein.de/unsere-partner).

The authors used Anthropic’s Claude to draft and edit prose and to write the analysis and verification scripts that recompute the reported values and figures from the archived simulation outputs.
The authors reviewed all such material and are responsible for its accuracy, integrity and originality.

\clearpage

\appendix

\section{Implementation and discretisation} \label{app:implementation}

\subsection{Solution procedure} \label{appsec:procedure}

The Cahn-Hilliard-Navier-Stokes equations are solved by the code \texttt{phaseFieldFoam} \cite{cai_numerical_2015} in a segregated manner \cite{jamshidi_suitability_2019, worner_spreading_2021}, using the cell-centred, collocated finite volume discretisation for arbitrary polyhedral meshes provided by foam-extend-4.1 \cite{jasak_release_2016, weller_tensorial_1998}.
The procedure for advancing the solution from time step $n\triangleq t^n$ to $n+1\triangleq t^{n+1}$ is as follows:
\begin{enumerate}
    \item Compute $\phi^n$ using $C^n$ according to \cref{eq:phi}.
    \item Solve the Cahn-Hilliard \cref{eq:CH} to obtain $C^{n+1}$ using $\phi^n$ and $\mathbf{u}^n$.
    \item Use $C^{n+1}$ to determine $\rho^{n+1}$ and $\mu^{n+1}$ by \cref{eq:rho,eq:mu}, $\phi^{n+1}$ by \cref{eq:phi}, and $\mathbf{
    f}_\sigma^{n+1}=-C^{n+1}\nabla\phi^{n+1}$ by \cref{eq:f_sigma}.
    \item Solve the Navier-Stokes \cref{eq:momentum,eq:continuity} to obtain $\mathbf{u}^{n+1}$ and $p^{n+1}$ using $\rho^{n+1}$, $\mu^{n+1}$, and $\mathbf{f}_\sigma^{n+1}$.
\end{enumerate}
Splitting the Cahn-Hilliard equation into \cref{eq:CH,eq:phi} and evaluating $\phi$ in a separate step, as in steps 1 and 2, avoids assembling the fourth-order operator that \cref{eq:CH} contains when $\phi$ is eliminated, so that only second-order operators have to be discretised.

\subsection{Discretisation schemes}  \label{appsec:schemes}
 
Time integration is performed with the second-order implicit \texttt{backward} scheme, and gradients are evaluated by Gauss integration with linear face interpolation and a cell limiter (\texttt{cellLimited Gauss linear 1}), which prevents values extrapolated from a cell centre to its faces from leaving the range of the neighbouring cell values \cite{jasak_error_1996}.

Convection of the phase field requires a scheme that is bounded, since $C$ is only meaningful within the range $-1\leq C\leq1$, and of low numerical diffusion, since the interface is resolved by only $N_C=8$ cells, cf.\ \cref{subsec:numericalParameters}, and would otherwise be smeared over a width set by the grid rather than by $\varepsilon$.
We therefore use the \texttt{Gamma} scheme \cite{jasak_high_1999} with a blending coefficient of $0.25$ (\texttt{Gauss Gamma 0.25}), which applies central differencing wherever the solution is locally smooth and blends smoothly towards upwind differencing only where it is not, so that boundedness is enforced without smearing the interface.
Momentum convection is discretised with a linear scheme whose limiter blends towards upwind differencing near steep gradients (\texttt{Gauss limitedLinearV 1}) \cite{van_leer_towards_1974}.
The \texttt{V} variant computes a single limiter for the velocity vector as a whole, rather than one per Cartesian component, so that the limiting does not introduce a spurious change in the direction of the interpolated velocity.
 
The remaining settings address the mesh quality.
The grids of \cref{sec:validation,sec:flatOrifices} are uniform and Cartesian, whereas the GDE-like structures of \cref{sec:meshes} are cut out of such a grid with \texttt{snappyHexMesh}, which leaves skewed and non-orthogonal cells along the solid surface, i.e., exactly where the contact line resides and where the interface shape is determined.
Face interpolations and convective fluxes therefore carry an explicit skewness correction (\texttt{skewCorrected}), and Laplacians and surface-normal gradients a non-orthogonal correction limited to at most half of the orthogonal contribution (\texttt{limited 0.333}), which keeps the explicitly treated part of the surface-normal gradient small against its implicit part on the poorest cells \cite{jasak_error_1996, moukalled_finite_2016}.
Both corrections are inactive on the Cartesian grids, so that an identical numerical setup is used throughout this work.

\section{Nomenclature} \label{app:nomenclature}

{\small
\setlength{\LTcapwidth}{\textwidth}
\begin{longtable}{@{}l >{\raggedright\arraybackslash}p{0.64\textwidth} l@{}}
\caption{Nomenclature. Symbols carrying a subscript $\mathrm{A}$ or $\mathrm{B}$ refer to the corresponding pure phase, cf.\ the list of subscripts below.}
\label{tab:nomenclature} \\
\hline
Symbol & Description & Unit \\
\hline
\endfirsthead
\hline
Symbol & Description & Unit \\
\hline
\endhead
\hline
\endfoot
\multicolumn{3}{@{}l}{\emph{Roman symbols}} \\
$A_\mathrm{s}$                  & open area of an opening                                     & \si{\square\meter} \\
$a$                             & side length of the projected mesh opening, $a=w-2r_\mathrm{w}$ & \si{\meter} \\
$a_\mathrm{eff}$                & effective aperture at the height of the contact line        & \si{\meter} \\
$AR$                            & aspect ratio of an opening, $AR=y_\mathrm{s}/x_\mathrm{s}$  & --- \\
$C$                             & order parameter (phase field), $C=\alpha_\mathrm{A}-\alpha_\mathrm{B}$ & --- \\
$d_\mathrm{s}$                  & equivalent orifice diameter, $d_\mathrm{s}=2r_\mathrm{s}$   & \si{\meter} \\
$f$                             & end-cap factor, \cref{eq:endCapFactor}                      & --- \\
$\mathbf{f}_\sigma$             & surface tension volume force                                & \si{\newton\per\cubic\meter} \\
$\mathbf{g}$                    & gravitational acceleration, $g=|\mathbf{g}|$ & \si{\meter\per\square\second} \\
$h$                             & interface height: of the apex above the orifice plane (\cref{sec:pressureCurve,sec:validation,sec:flatOrifices}), or at the centre of the opening above the mesh mid-plane (\cref{sec:meshes}) & \si{\meter} \\
$h_0$                           & interface height at $\Delta p=0$ (pressure-balanced state)  & \si{\meter} \\
$h_\mathrm{res}$                & reservoir height                                            & \si{\meter} \\
$h_\mathrm{s}$                  & orifice height                                              & \si{\meter} \\
$K$                             & interface stiffness, $\mathrm{d}\Delta p/\mathrm{d}h$ at the centre of the admissible window, \cref{eq:stiffness} & \si{\pascal\per\meter} \\
$L_C$                           & width of the diffuse interface, $L_C=4.164\varepsilon$      & \si{\meter} \\
$L_\mathrm{c}$                  & characteristic macroscopic length scale                     & \si{\meter} \\
$\ell$                          & grid cell size                                              & \si{\meter} \\
$N_C$                           & number of cells resolving the diffuse interface             & --- \\
$\mathbf{n}$                    & unit normal vector of a boundary                            & --- \\
$\mathbf{n}_\mathrm{s}$         & unit normal pointing from the fluid into the solid          & --- \\
$p$                             & pressure                                                    & \si{\pascal} \\
$\Delta p$                      & Laplace pressure across the interface, gas minus liquid & \si{\pascal} \\
$\Delta p_\mathrm{max}$         & maximum admissible pressure difference (gas breakthrough)   & \si{\pascal} \\
$\Delta p_\mathrm{min}$         & minimum admissible pressure difference (flooding); zero for a pinned flat orifice & \si{\pascal} \\
$\Delta p_\mathrm{w}$           & width of the operating window, \cref{eq:windowWidth}        & \si{\pascal} \\
$\Delta p_\mathrm{s}$           & analytical Laplace pressure at the orifice plane, \cref{eq:dp_orifice} & \si{\pascal} \\
$R_0$                           & radius of curvature at the interface apex                   & \si{\meter} \\
$R_1$, $R_2$                    & principal radii of curvature                                & \si{\meter} \\
$R_\mathrm{min}$                & smallest attainable mean radius of curvature                & \si{\meter} \\
$r_\mathrm{res}$                & reservoir radius                                            & \si{\meter} \\
$r_\mathrm{s}$                  & radius of the area-equivalent circular orifice              & \si{\meter} \\
$r_\mathrm{w}$                  & wire radius                                                 & \si{\meter} \\
$s$                             & arc length along the interface profile                      & \si{\meter} \\
$t$                             & time                                                        & \si{\second} \\
$\Delta t$                      & time step width                                             & \si{\second} \\
$\Delta t_\mathrm{max}$         & upper bound of the time step width & \si{\second} \\
$\mathbf{u}$                    & velocity field                                              & \si{\meter\per\second} \\
$U_0$                           & inflow velocity driving the interface                       & \si{\meter\per\second} \\
$U_\mathrm{max}$                & maximum velocity magnitude in the domain                    & \si{\meter\per\second} \\
$w$                             & wire pitch of the woven mesh                                & \si{\meter} \\
$x$, $y$, $z$                   & Cartesian coordinates                                       & \si{\meter} \\
$\tilde{x}$, $\tilde{z}$        & radial distance and depth measured from the bubble apex     & \si{\meter} \\
$x_\mathrm{s}$, $y_\mathrm{s}$  & width and length of an opening; for a diamond, its short and long diagonal & \si{\meter} \\
$z_0$                           & initial position of the flat interface                      & \si{\meter} \\
$z_\mathrm{res,min}$, $z_\mathrm{res,max}$& lower and upper boundary of the mesh domain, \cref{subsec:meshSetup} & \si{\meter} \\
\multicolumn{3}{@{}l}{} \\
\multicolumn{3}{@{}l}{\emph{Greek symbols}} \\
$\alpha$                        & opening angle of a diamond shaped orifice                   & \si{\degree} \\
$\alpha_\mathrm{A}$, $\alpha_\mathrm{B}$ & volumetric phase fractions                         & --- \\
$\delta h$, $\delta p$          & interface displacement and pressure fluctuation             & \si{\meter}, \si{\pascal} \\
$\varepsilon$                   & interfacial width parameter                                 & \si{\meter} \\
$\theta_\mathrm{e}$             & equilibrium (static) contact angle, measured in phase A     & \si{\degree} \\
$\kappa$                        & mobility, $\kappa=\chi\varepsilon^2$                        & \si{\cubic\meter\second\per\kilogram} \\
$\lambda$                       & mixing energy density                                       & \si{\newton} \\
$\mu$                           & dynamic viscosity; $\mu_C$ local value, \cref{eq:mu} & \si{\pascal\second} \\
$\rho$                          & density; $\rho_C$ local value, \cref{eq:rho} & \si{\kilogram\per\cubic\meter} \\
$\sigma$                        & surface tension                                             & \si{\newton\per\meter} \\
$\varphi$                       & angle between the interface tangent and the horizontal      & \si{\degree} \\
$\phi$                          & chemical potential                                          & \si{\pascal} \\
$\chi$                          & mobility scaling constant                                   & \si{\meter\second\per\kilogram} \\
\multicolumn{3}{@{}l}{} \\
\multicolumn{3}{@{}l}{\emph{Dimensionless numbers}} \\
$Bo$   & Bond number, $Bo=(\rho_\mathrm{A}-\rho_\mathrm{B})gr_\mathrm{s}^2/\sigma$ & --- \\
$Ca$                            & capillary number, $Ca=\mu_\mathrm{A}U_0/\sigma$ & --- \\
$Cn$                            & Cahn number, $Cn=\varepsilon/L_\mathrm{c}$, here with $L_\mathrm{c}=d_\mathrm{s}$ & --- \\
$Co$   & Courant number, $Co=U_\mathrm{max}\Delta t/\ell$                          & --- \\
$Co_\mathrm{max}$               & upper bound of the Courant number & --- \\
$We$                            & Weber number, $We=\rho_\mathrm{A}U_0^2 r_\mathrm{s}/\sigma$ & --- \\
\multicolumn{3}{@{}l}{} \\
\multicolumn{3}{@{}l}{\emph{Subscripts}} \\
$\mathrm{A}$, $\mathrm{B}$      & phase A (water) and phase B (air), also written $\mathrm{water}$, $\mathrm{air}$ &  \\
$C$                             & depending on the order parameter ($\rho_C$, $\mu_C$), or of the diffuse interface ($L_C$, $N_C$) &  \\
$\mathrm{c}$                    & characteristic ($L_\mathrm{c}$) &  \\
$\mathrm{e}$               & equilibrium                        & \\
$\mathrm{eff}$             & effective                          & \\
$\mathrm{max}$, $\mathrm{min}$ & maximum, minimum               & \\
$\mathrm{rect}$, $\mathrm{slit}$ & rectangular orifice, slit    & \\
$\mathrm{res}$             & reservoir                          & \\
$\mathrm{s}$               & solid structure (orifice or opening) & \\
$\mathrm{w}$               & wire ($r_\mathrm{w}$) or operating window ($\Delta p_\mathrm{w}$) & \\
$0$                             & apex ($R_0$), pressure-balanced state ($h_0$), initial or inflow value ($z_0$, $U_0$) &  \\
\end{longtable}
}

\bibliographystyle{elsarticle-num}
\bibliography{BIB.bib}

\end{document}